%% file: main.tex
\documentclass[final,5p,times,twocolumn]{elsarticle}
\makeatletter
\def\ps@pprintTitle{%
   \let\@oddhead\@empty
   \let\@evenhead\@empty
   \let\@oddfoot\@empty
   \let\@evenfoot\@empty
}
\makeatother

\usepackage{amssymb}
\usepackage{lipsum}
\usepackage{amsthm}

\usepackage[normalem]{ulem}
\usepackage{graphicx,subfigure}
\usepackage{float}
\usepackage{dcolumn}
\usepackage{cuted}
\usepackage{bm}
\usepackage{pdfpages}
\usepackage{epstopdf}
\usepackage{mathrsfs}
\usepackage{amssymb,amsmath,amsfonts,latexsym}
\usepackage{nicefrac}
\usepackage[colorlinks=true,linktocpage=true,linkcolor=blue,citecolor=blue]{hyperref}
\usepackage[utf8]{inputenc}
\usepackage[T1]{fontenc}
\usepackage{natbib}
\usepackage{lipsum}
\usepackage{nicefrac}
\usepackage{slashed}
\usepackage{mathtools}
\usepackage{multirow}
\usepackage{caption}
\usepackage{lineno}
\usepackage{orcidlink}

\input{definitions.tex}

\begin{document}
\begin{frontmatter}
\title{Predictions for photon-jet correlations at forward rapidities in heavy-ion collisions}
\author[a,b]{Souvik Priyam Adhya\orcidlink{https://orcid.org/0000-0002-4825-2827}}
\ead{adhya@fzu.cz}
\author[a]{Krzysztof Kutak\orcidlink{https://orcid.org/0000-0003-1924-7372}}
\ead{krzysztof.kutak@ifj.edu.pl}
\author[c]{Wies{\l}aw P{\l}aczek\orcidlink{https://orcid.org/0000-0002-9678-9303}}
\ead{wieslaw.placzek@uj.edu.pl }
\author[d,e]{Martin Rohrmoser\orcidlink{https://orcid.org/0000-0003-2311-832X}}
\ead{rohrmoser.martin1987@gmail.com }
\author[f]{Konrad Tywoniuk\orcidlink{https://orcid.org/0000-0001-5677-0010}}
\ead{konrad.tywoniuk@uib.no}
\affiliation[a]{Institute of Nuclear Physics, Polish Academy of Sciences,
ul. Radzikowskiego 152, 31-342 Krakow, Poland}
\affiliation[b]{Institute of Physics of the Czech Academy of Sciences, Na Slovance 1999, 18221 Prague 8, Czech Republic}
\affiliation[c]{Faculty of Physics, Astronomy and Applied Computer Science and Mark Kac Center for Complex Systems Research, Jagiellonian University, ul. Lojasiewicza 11, 30-348 Krakow,  Poland}
\affiliation[d]{Faculty of Material Engineering and Physics, Cracow University of Technology, ul. Podchorazych 1, 30-084 Krakow, Poland}
\affiliation[e]{Institute of Physics, Jan Kochanowski University, ulica Uniwersytecka 7, P-25-406 Kielce, Poland}
\affiliation[f]{Department of Physics and Technology, University of Bergen, 5007 Bergen, Norway}
\begin{abstract}
In this work, we study for the first time jet-medium interactions in heavy-ion collisions with introduction of saturation and Sudakov effects with parameters tuned for upcoming forward calorimeter acceptances in experiments, in particular the ALICE FoCal detector. We focus on $\gamma +$jet correlations by taking into account in-medium parton evolution using the BDIM equation that describes jet interactions with the quark-gluon plasma (QGP) combined with vacuum-like emissions (VLE). We systematically introduce the early time gluon saturation dynamics through the small-$x$ Improved Transverse Momentum Dependent factorization (ITMD). For our purpose, we use Monte Carlo programs \katie\ and \tmdice\ to generate hard events and in-medium parton evolution, respectively. We present results of azimuthal correlations and nuclear modification ratios to gauge the impact of the gluon saturation effects at early time for the in-medium jet energy loss.  
\end{abstract}
\end{frontmatter}
\section{Introduction}\label{introduction}
Quantum Chromodynamics (QCD)  is a well established theory of strong interactions. However, it predicts phenomena that still need experimental confirmation. One such a puzzle is the phenomenon of ``gluon saturation'' \cite{Gribov:1984tu}, where the growth of gluon density slows down due to a balance between gluon splittings and recombinations. The theoretical formalism that is developed to tackle saturation is the Color Glass Condensate (CGC) effective theory (as reviewed in \cite{Gelis:2010nm, Kovchegov:2012mbw}). Saturation is a QCD prediction anticipated at high-energy hadron collisions. While multiple experimental indications suggest its presence -- exemplified by direct data \cite{STAR:2006dgg, PHENIX:2011puq, STAR:2021fgw} as well as saturation-based approaches correlated with the HERA, RHIC and LHC data \cite{Golec-Biernat:1998zce, Lappi:2012nh,Ducloue:2017kkq, Albacete:2018ruq, Ducloue:2016ywt,Stasto:2018rci, vanHameren:2019ysa, Albacete:2018ruq,Benic:2022ixp}.

The LHC program offers significant potential for uncovering evidence of the gluon saturation. However, currently, no LHC experiments explicitly aim to study saturation physics, most likely to be observed in processes where longitudinal momentum of a target is probed at values $x<10^{-5}$. In the LHC kinematics for jets this corresponds to particle production in a forward rapidity region. The forthcoming FoCal calorimeter of the ALICE collaboration will bridge this gap \cite{ALICE:2023fov, ALICE:2020iug}. Designed to achieve high-resolution measurements of jets and photons in the forward region, it will cover pseudorapidities within the range $3.4 < \eta < 5.8$. These possible measurements have caught a lot of attention and motivated many predictions which are optimistic for saturation to be found in $p$A collisions as they offer nuclear density enhancement in favor of saturation as compared to $pp$ collisions \cite{Dumitru:2005gt,Marquet:2007vb,vanHameren:2016ftb,Stasto:2018rci,Jucha:2023hjg,Ganguli:2023joy,vanHameren:2023oiq,Altinoluk:2021ygv,Fujii:2020bkl} (see also \cite{Morreale:2021pnn} for review).

An intriguing possibility for investigating saturation is to study forward jet production in AA collisions. This system, similarly to $p$A collisions, offers enhancement of nuclear density. The process is, however, more complex, as scattering events of heavy-ions involving large momentum transfer  result in creation of highly energetic jets immersed in a medium. These jets serve as ``probes'' for examining the properties of the densely interacting medium formed during high-energy heavy-ion collisions~\cite{Shuryak:1978ij,Shuryak:2004cy}. 
The basic goal of the paper is to investigate to what extend the saturation effects are visible in final states of heavy-ion collisions. 
Naively, one could think that the saturation effects will be lost in the medium  because the quark-gluon plasma
(QGP) medium strongly modifies the properties of jets. 
However, the point is that the Forward Calorimeter (FoCal) of the ALICE experiment offers a possibility to look specifically in the forward rapidity region,  
that allows to probe parton densities at a very low $x$ region, where in $p$Pb collisions the saturation effects were shown to be large and where back-to-back configurations are suppressed \cite{Marquet:2007vb,Kutak:2012rf,vanHameren:2023oiq}. 
In addition, both the ATLAS and the CMS detectors will, in the future, also feature a Forward Calorimeter (FCal) and a Hadronic Forward (HF) calorimeter with the pseudorapidity ranges of $3.1 < \eta < 4.9$ and $3.0 < \eta < 5.2$, respectively \cite{Aad:2008zzm, Chatrchyan:2008aa}, crucial for capturing and studying forward-scattering events. 
We therefore expect that the suppression of particle production in the forward acceptance region of the detector will be large enough to be visible in PbPb collisions.

The production of high-energy particles, like partons or photons, in ``hard'' processes occurs over very short time intervals, with $\tau\sim 1/\pt \lesssim 0.1$ fm/c. Consequently, their 
features
can potentially be altered by interactions in the final state while they traverse the medium. The cross-sections for energetic particle production can be calculated using perturbative QCD methods, making them valuable `tomographic' tools for probing the properties of the created medium \cite{Appel:1985dq,Blaizot:1986ma,Gyulassy:1990ye,Wang:1991xy,Baier:1996sk,Zakharov:1997uu,Neufeld:2010fj,Adhya:2022tcn,Adhya:2024nwx}.

In collisions involving Pb nuclei at the LHC, the impact of the produced medium has been investigated at mid-rapidity by studying pairs of back-to-back dijets. These dijets displayed significant imbalance in their transverse momenta \cite{Chatrchyan:2011sx,Aad:2010bu}. 
However, the advantage of having a large yield of dijets compared to pairs of $\gamma$+jet is counterbalanced by the loss of information about the initial characteristics of the probes, specifically prior to their interactions with the medium. Additionally, correlating two probes that both undergo energy loss introduces a bias toward scatterings happening at the medium surface and aligned tangentially to it. At the
leading order (LO) in the QCD coupling constant, photons ($\gamma$) are generated in tandem with a closely associated quark (jet) having the same transverse momentum. Furthermore, these photons interact only weakly with the medium. The yield of isolated photons in PbPb collisions was found to align with expectations based on $pp$ data and the number of nucleon--nucleon collisions, with a nuclear modification factor $R_{AA} = 0.99\pm 0.31\;\text{(stat.)}\pm 0.26\;\text{(syst.)}$. As a result, the production of \photonjet\ combinations%
\footnote{The \photonjet\ cross section measured in $pp$ as well as PbPb collisions gets contributions from both `prompt' and `fragmentation' photons. The former denotes photons that are directly produced in the hard sub-processes, whereas the latter 
originates from the fragmentation of a jet involved in the $2\to 2$ hard matrix element. In the present work, we consider only the contribution from the prompt photons in the forward region.} 
has been identified as the optimal means for investigating the energy loss of partons within the medium \cite{Wang:1996yh,Wang:1996pe}, recently through parametric modeling approach in central rapidities \cite{Ogrodnik:2024qug}. However, all these studies were performed in the central rapidity regions where the impact of the gluon saturation physics is expected to be small. In experimental contexts, events with an increased yield of prompt photons are selected using an isolation requirement. 
This requirement dictates that additional energy within a fixed-radius cone around the reconstructed photon direction must fall below a specified threshold. Consequently, ``isolated photons'' are primarily composed of prompt photons generated directly during the initial hard scattering. This isolation condition is used to suppress background photons originating from decays of neutral mesons, such as $\pi_0$, $\eta$ and $\omega$, which are predominantly created through jet fragmentation.

The paper is organized as follows. Section~\ref{sec-ITMD} introduces the gluon saturation physics as the initial conditions in the hard processes of the \photonjet\ formation. Next, in Section~\ref{sec:inmedium}, we describe the energy loss mechanism of the jet inside the medium undergoing quenching due to vacuum-like emissions as well as coherent energy loss and multiple scatterings within the medium. In Section~\ref{sec:numresults}, we present the results for the magnitude of the jet quenching in the forward rapidity region based on anticipated parameters of the FoCal detector. Additionally, we present quantitative comparisons between different configurations of the saturation and medium-jet suppression effects for rapidity and azimuthal angular distributions. Finally, in Section~\ref{sec:conclusions}, we summarize our findings and present conclusions.
In Appendix~A, we discuss some uncertainties of our analyses, and in Appendix~B, we describe our Monte Carlo algorithm used for generation of vacuum-like emissions (VLE).

\section{Pre-medium scenario}
\label{sec-ITMD}
%
\begin{figure}[ht!]
    \centering
    \includegraphics[width=0.48\textwidth]{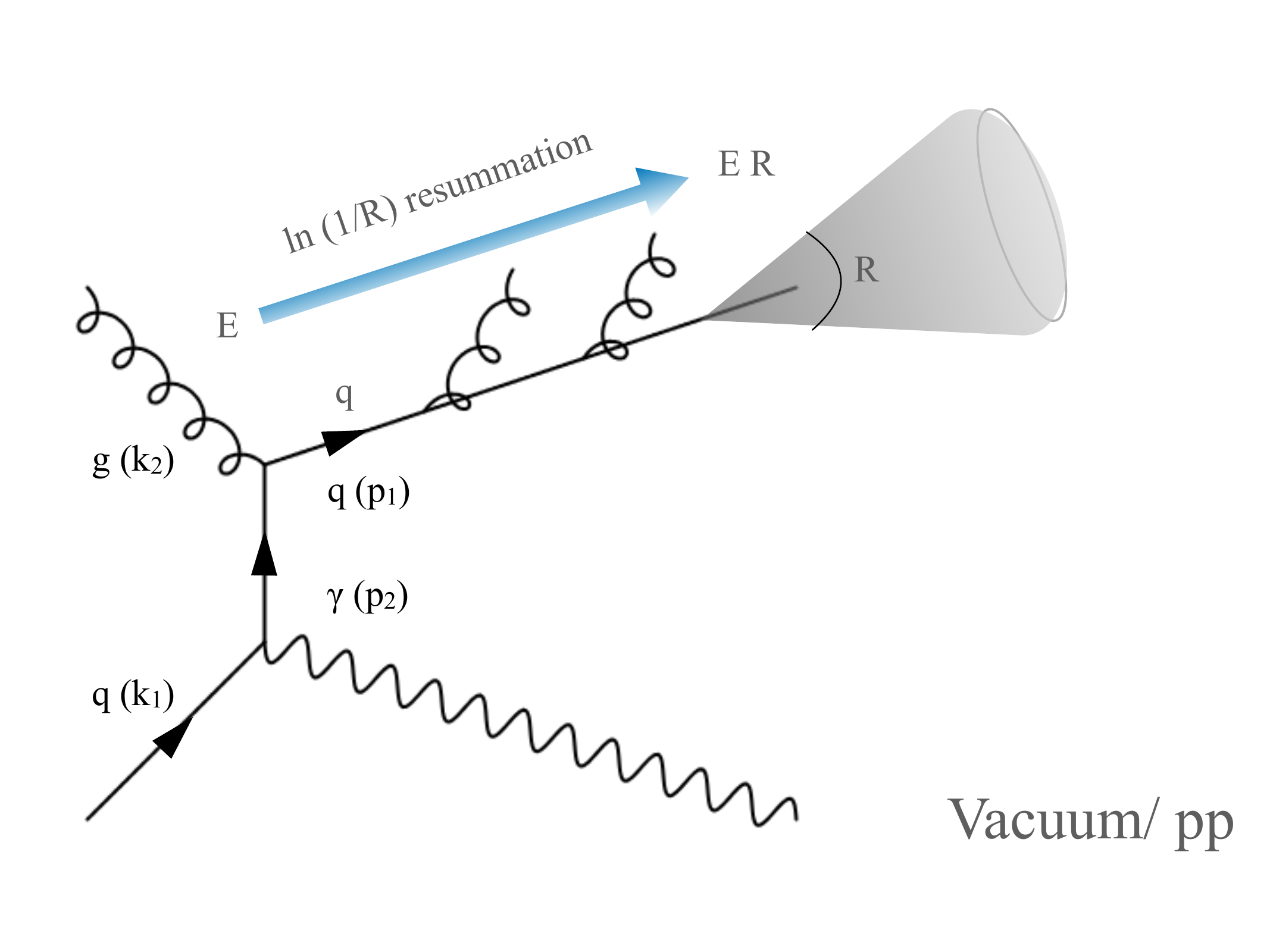}\hspace{-0.5cm}
    \includegraphics[width=0.48\textwidth]{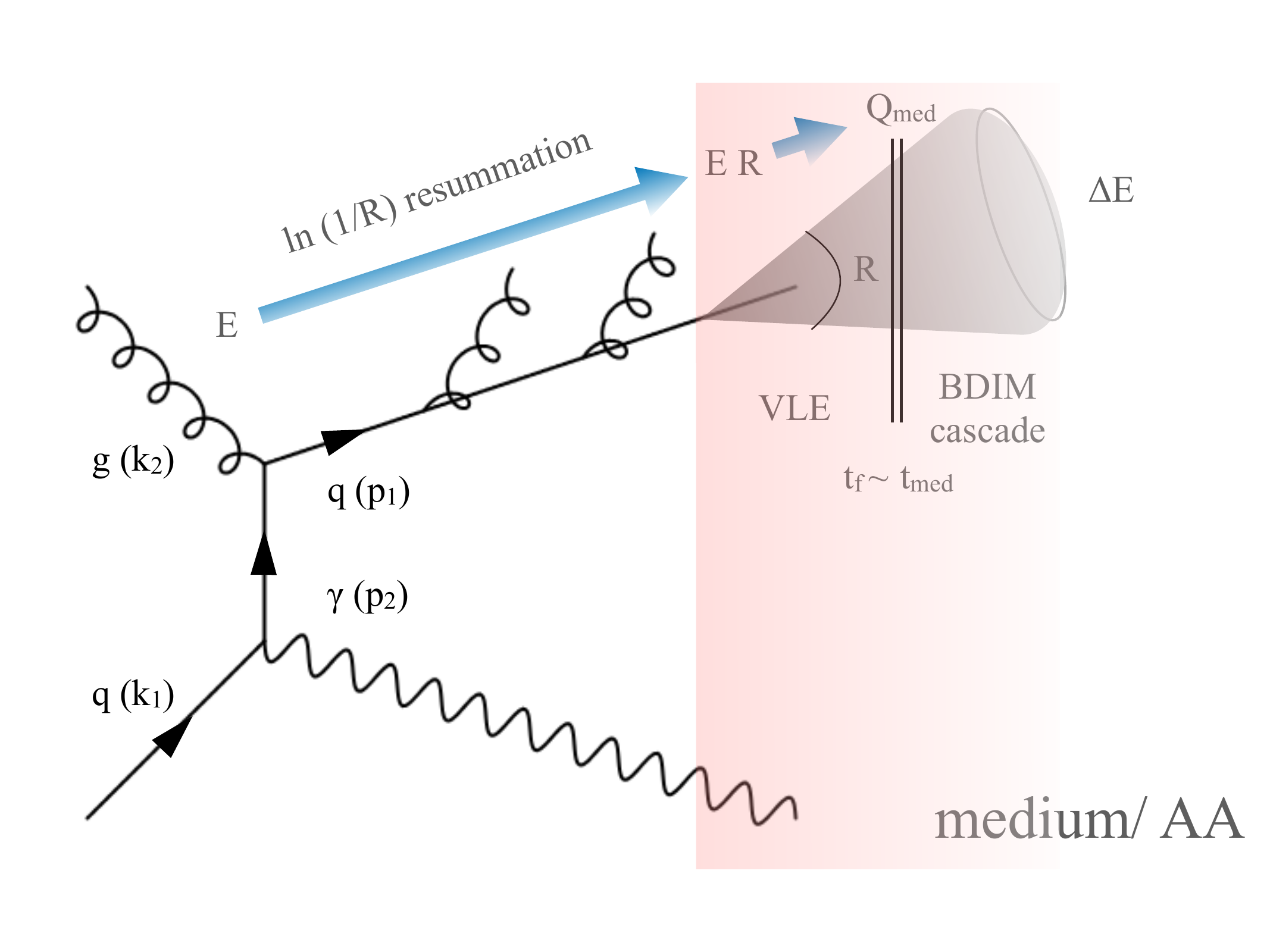}
    \caption{Quark-jet and prompt photon production via  scattering of a quark and a gluon from hard nuclear collisions. The propagation of the jet for the vacuum /$pp$ scenario (top) and medium/AA scenario (bottom), respectively.} 
    \label{fig:vac_pp-med_AA}
\end{figure}

As mentioned in the previous section, the process we consider here, depicted schematically in Fig.~\ref{fig:vac_pp-med_AA}, 
involves $\gamma +$jet final states being produced 
in the forward direction in high-energy heavy-ion collisions:
\begin{equation}
A_1 + A_2\rightarrow \gamma + \mathrm{jet} +X\,,
\label{eq:collision}
\end{equation}
where $A_1, A_2$ denote two colliding ions (e.g.\ of Pb) and $X$ is an unmeasured final state.
The effective theory that allows to calculate the final states in this scenario is CGC, as already mentioned in Introduction. It can be further shown that it leads to the so-called Improved Transverse Momentum Dependent Factorization (ITMD) \cite{Kotko:2015ura,Altinoluk:2019fui} which is formulated in the momentum space. The later framework is expressed as gauge-invariant transverse-momentum-dependent (TMD) parton density functions convoluted with a gauge-invariant hard matrix element. For a generic process, this framework represents a limit of CGC neglecting genuine twists, i.e.\ contributions where $Q_s\!<\!p_T$. This is a good approximation when considering jet physics. For the process that we consider the relation is exact, i.e. ITMD is equivalent to CGC \cite{Rezaeian:2012wa,Jalilian-Marian:2012wwi}.

The leading pre-quenching partonic process is
\begin{equation}
q(k_1) + g^*(k_2)\rightarrow  q(p_1) + \gamma(p_2)\,,
\label{eq:parproc}
\end{equation}
where $q$ denotes on-shell quarks while $g^*$ -- a space-like gluon.
In the configuration where the final state is such that the observed jet and photon are both in the forward direction, the longitudinal momenta of the incoming partons%
\footnote{The momentum fractions $x_i$, the rapidities $y_i$ and the transverse momenta $q_{iT}$ of the outgoing particles are related to one another as: $x_1 = [q_{1T}\rme^{y_1}  + q_{2T} \rme^{y_2}]/\sqrt{s}$ and $x_2 = [q_{1T}\rme^{-y_1}+ q_{2T}\rme^{-y_2}]/\sqrt{s}$.} are strongly ordered \cite{Dumitru:2005gt,Marquet:2007vb,Deak:2009xt,Kutak:2012rf}: 
\begin{align}
k_1=x_1\,P_1,&& k_2=x_2\,P_2+k_{2T}\,, 
\end{align}
with $x_1\gg x_2$.

The cross section for  the pre-quenching process is a direct extension of a cross-section formula for $pp$ and $p$A collisions, 
and it reads \cite{Deak:2009xt,Kutak:2012rf} 
\begin{align}
 \frac{d\sigma^{AA\rightarrow \gamma +\mathrm{jet}+X}}
       {dy_1dy_2dp_{1T}dp_{2T}d\Delta\phi}
 &=\frac{p_{1T}p_{2T}}{8\pi^2 (x_1x_2 s)^2}
  \sum_{a} 
  x_1 f_{a/A}(x_1,\mu_F^2)\,\nn
 & \times |{\cal M}_{ag^*\to \gamma a}^{\mathrm{off-shell}}|^2
  {\cal F}(x_2,k_{2T}^2,\mu_F^2)\,,
  \label{eq:hef-formula}
\end{align}
where ${\cal M}^{\mathrm{off-shell}}_{a g^* \to \gamma a}$ is the matrix element with one off-shell and one on-shell parton in the hard sub-process, $f_{a/A}(x_1,\mu^2)$ is the collinear quark-density distribution for the lead nucleus and ${\cal F}(x_i,k_{iT}^2,\mu_{F}^2)$ is the operator-defined dipole gluon density obeying the nonlinear Balitsky--Kovchegov equation \cite{Balitsky:1995ub,Kovchegov:1999yj}. The nonlinearity of the equation allows for recombination of gluons, which in the end leads to emergence of a saturation scale.
The dipole gluon density, besides its dependence on $x$ and $k_T$, it also depends on the hard scale introduced via the Sudakov form factor \cite{Mueller:2013wwa,Mueller:2012uf,Al-Mashad:2022zbq}.
The Sudakov form factor that we use is appropriate for the $\qg^*\to\gamma q$ hard process and it reads
\begin{equation}
S(\mu_F,b_\perp)=-\frac{\alpha_s}{2\pi}\left( \frac{N_c}{2}+\frac{C_F}{2}\right)\ln^2\frac{\mu_F^2b_\perp^2}{c_0^2} ,
\end{equation}
where $b_\perp$ is a Fourier conjugate scale to $k_T$, $\mu$ is a hard scale and $c_0=2e^{-\gamma_E}$ ($\gamma_E$ is the Euler constant); 
more details of its implementation can be found in Ref.~\cite{Al-Mashad:2022zbq}.
\section{Medium scenario}\label{sec:inmedium}


The evolution of parton jets produced in heavy-ion collisions undergoes further emissions, see Fig.~\ref{fig:vac_pp-med_AA}, based on characteristic timescales of the processes%
\footnote{After jet particles leave the medium, they can still fragment via emissions of partons. We neglected so far a possible third stage of jet evolution \cite{Caucal:2018dla,Mehtar-Tani:2012mfa} where the jet particles undergo further branching in the vacuum after leaving the QGP medium.}.
\begin{enumerate}
    \item At the beginning of jet evolution the color degrees of freedom of individual jet particles may not be resolved by the particles within a QGP medium. However, parton radiation in the vacuum, known as the vacuum-like emissions (VLE), is still possible \cite{Caucal:2018dla}. Therefore, the evolution via such a parton emission follows the Dokshitzer--Gribov--Lipatov--Altarelli--Parisi (DGLAP) evolution equations \cite{Gribov:1972ri, Altarelli:1977zs, Dokshitzer:1977sg}. Phase-space boundaries at which the colors of individual jet particles are resolved by the medium are described in Subsection~\ref{sec:vle}.
    \item  As the medium particles resolve the colors of the individual jet particles, the latter undergo processes of scattering off the former as well as the medium-induced radiation. Here, we implement the evolution that follows the equations by Blaizot--Dominguez--Iancu--Mehtar-Tani (BDIM)~\cite{Blaizot:2012fh,Blaizot:2013vha,Blanco:2021usa}.
    We refer to this stage of the process as the BDIM evolution and discuss it in more detail in Subsection~\ref{sec:medium}.
\end{enumerate}

\subsection{Vacuum-like emissions}
\label{sec:vle}

In a vacuum, the high-energy partons produced as a result of the hard QCD collision undergo a reduction in their large virtuality down to the hadronization scale through successive splittings. These processes have been studied in detail theoretically \cite{Dasgupta:2014yra, Dasgupta:2016bnd, Kang:2016mcy} and implemented within Monte Carlo (MC) parton shower generators through the DGLAP evolution equations. It has been firmly established that there is a significant phase space for VLE occurring before any medium-induced modifications can develop \cite{Caucal:2018dla,Mehtar-Tani:2017web,Dominguez:2019ges,Mehtar-Tani:2024jtd}. This vacuum-like phase space significantly impacts the amount of jet-energy loss based on the vacuum-set scales governing jet activity, such as multiplicity. Therefore, understanding this phase space is crucial for explaining numerous jet-quenching observables, such as the relative suppression between jets and hadrons \cite{Mehtar-Tani:2017web, Casalderrey-Solana:2018wrw, Casalderrey-Solana:2019ubu, Caucal:2019uvr}. In light of these recent developments, we consider the possibility that a particle created in a hard initial collision can radiate partons as in the vacuum before entering a QGP-medium.
Therefore, we presume that the initial parton $i$ undergoes parton branchings according to probability densities 
for emitting a particle $j$ given as
\begin{equation}
\frac{\rmd^2\mathcal{P}_{ji}}{\rmd Q^2\rmd z}=\frac{\alpha_s}{2\pi}\frac{1}{Q^2}P_{ji}(z)\,,
\label{eq:vlerate}
\end{equation}
where $Q$ is the virtuality of the branching particle, $z$ is the energy fraction of the emitted particle and the functions $P_{ji}(z)$ are the DGLAP splitting functions at LO in QCD, given as
\begin{align}
P_{qq}(z)&=C_F\frac{1+z^2}{1-z}\,,\\
P_{gq}(z)&=P_{qq}(1-z)\,,\\
P_{qg}(z)&=T_R \left[z^2+(1-z)^2 \right]\,,\\
P_{gg}(z)&=C_A \left[\frac{1-z}{z}+\frac{z}{1-z}+z(1-z)\right]\,.
\end{align}
As shown in~\cite{Caucal:2018dla,Mehtar-Tani:2012mfa}, 
 after entering the QGP medium, the individual color degrees of freedom of two jet-particles $i$ and $j$ may not be resolved by the medium for a certain decoherence time ($t_{\rm decoh}$) that can be estimated as~\cite{Mehtar-Tani:2011hma,Adhya:2021kws}
\begin{equation}
t_{\rm decoh}\approx \left(\frac{12}{\hat{q} \theta_{ij}^2}\right)^{1/3}\,,
\label{eq:decoherencetime}
\end{equation}
where $\theta_{ij}$ is the branching angle for which a small angle approximation
\begin{equation}
    \theta_{ij}\approx Q_i/\sqrt{E_i^2z(1-z)}\,,
\end{equation}
was used. The average transverse momentum transfer $\hat{q}$ is defined via the changes over time $t$ of a particle traveling through the medium with the momentum component $k_\perp$ transverse to its original incident direction as
\begin{equation}
\hat{q}=\frac{\rmd\langle k_\perp^2\rangle}{\rmd t} \,.
\end{equation}

On the other hand, a branching time of a radiating jet-particle $i$ due to the formation of bremsstrahlung can be estimated as 
\begin{equation}
t_{\rm br}\approx 2 E_i/Q_i^2\,.
\label{eq:branchingtime}
\end{equation}
Thus, one can estimate~\cite{Caucal:2018dla,Adhya:2021kws,Mehtar-Tani:2011lic,Mehtar-Tani:2012mfa} 
that as long as a pair of jet partons fulfills the condition
\begin{equation}
    t_{\rm br}<t_{\rm decoh}\,,
    \label{eq:stopcoh}
\end{equation}
parton branchings due to bremsstrahlung as in the vacuum occur faster than interactions with medium particles.
Therefore, we consider emissions following Eq.~(\ref{eq:vlerate}) for jet particles in the QGP medium as long as Eq.~(\ref{eq:stopcoh}) is not violated. 
Afterwards, both scatterings off medium particles as well as coherent medium-induced radiations that follow the well-known spectrum found by Baier, Dokshitzer, Mueller, Peign\'e and Schiff, and independently by Zakharow (BDMPS-Z)~\cite{Baier:1994bd,Baier:1996vi,Baier:2000mf,Baier:2000sb,Zakharov:1996fv,Zakharov:1997uu,Zakharov:1999zk} occur. It is to be noted that an additional constraint for VLEs is given by
\begin{equation}
    t_{\rm decoh}<t_{\rm L},
    \label{eq:stoplength}
\end{equation} 
where $t_{\rm L}$ is the time at which the jet particles leave the medium. This constraint implements the requirement that the splitting is resolved within a finite distance inside the medium. This is equivalent to requiring that the splitting has an angle greater than a critical angle $\theta_{ij} > \theta_c$, where $\theta_c = \sqrt{12/(\hat q L^3)}$. Using the double logarithmic approximation (DLA), where the multiplicity of gluons is dominated by the soft and collinear divergences in QCD, i.e. $\rmd N_\text{\tiny DLA} = \bar \alpha (\rmd z/z)(\rmd \theta/\theta)$ with $\bar \alpha = \alpha_s C_i/\pi$, the number of emitted gluons inside the medium can be estimated as \cite{Mehtar-Tani:2017web} 
\begin{equation}
\label{eq:in-phase-space}
N_\text{\tiny DLA}^\text{in} = 2 \bar \alpha \ln \frac{R}{\theta_c} \left(\ln\frac{E}{\omega_c} + \frac{2}{3} \ln \frac{R}{\theta_c} \right) \,,
\end{equation}
where $R$ is the jet-cone size and $\omega_c = \hat q L^2/2$ is the medium energy scale from multiple scattering.

It is  important to distinguish here the longitudinal energies $E$ and $\omega$ of the leading and sub-leading jet particles, respectively, and the transverse momentum $\pT$ as measured in the detector. When measured at mid-rapidity, the jets are propagating perpendicularly to the beam axis and we can identify $E = \pT$. At forward rapidities, $\eta>1$, a very rough estimate gives $E \sim \pT \rme^{\eta}$. Comparing to Eq.~\eqref{eq:in-phase-space}, we see that every unit of rapidity contributes for $E>\omega_c$, that is with an additional gluon emitted deep inside the QGP. Studying jets at forward rapidities therefore places more importance to details of the in-medium jet fragmentation and the multi-parton quenching.

The jet-evolution via VLEs is implemented by a Markov Chain Monte Carlo (MCMC) algorithm. 
In its current form, the algorithm simulates time-like parton cascades by selecting parton branchings with energy fractions $x$ and decreasing parton virtualities $Q$ of the emitter.
It is assumed that the emissions happen inside a cone of the opening angle $R$, which imposes an additional constraint on the selection of $x$ and $Q$.
\subsection{In-medium jet-evolution}\label{sec:medium}
After the jet has evolved via VLE, as described in the previous section, its individual color degrees of freedom can be resolved by medium particles allowing for processes of scattering and induced radiation. 
For sufficiently high particle energies and densities of scatterers, coherent emissions that follow the well-known BDMPS-Z spectra of Refs.~\cite{Baier:1996kr, Baier:1996sk, Zakharov:1996fv, Zakharov:1997uu} occur.
Therefore, we consider a coherent branching of a parton $i$ into partons $j$ and $k$ to occur with the probability density ~\cite{Blaizot:2012fh,Blaizot:2013vha,Blanco:2021usa}
\begin{equation}
    \frac{\rmd^4 \mathcal{P}_{ji}}{\rmd t\, \rmd z\, \rmd^2 \mathbf{Q} }=\frac{\alpha_s}{(2\pi)^2}\,\mathcal{K}_{ji} (\mathbf{Q},z, k_{i+})
    \label{eq:splitrate}
\end{equation}
at a time $t$, where $z$ is the light-cone energy fraction defined as $z=k_{j+}/k_{i+}$
with the light cone energies $k_{i+}$ and $k_{j+}$ of particles $i$ and $j$, respectively.  In Eq.~(\ref{eq:splitrate}), $\mathbf{Q}$ is given by the transfer in momentum components of particles $i$, $j$ and $k$ transverse to the jet axis $\mathbf{k}_i$, $\mathbf{k}_j$ and $\mathbf{k}_k$, respectively, as
\begin{align}
    \mathbf{k}_j&=z\mathbf{k}_i+\mathbf{Q}\,,\nonumber\\
    \mathbf{k}_k&=(1-z)\mathbf{k}_i-\mathbf{Q}\,.
\end{align}

The splitting kernel in Eq.~(\ref{eq:splitrate}) for a static medium profile is given as \cite{Baier:1996kr, Baier:1996sk, Zakharov:1996fv, Zakharov:1997uu}%
\footnote{In principle, one can implement a generic expanding scenario for 1D longitudinally Bjorken-expanding medium kernels, see e.g. \cite{Adhya:2019qse, Adhya:2021kws}.} 
\begin{equation}
{\cal K}_{ji}(\mathbf{Q},z,p_+)=\frac{2}{p_+}\frac{P_{ji}(z)}{z(1-z)}\sin\left[\frac{Q^2}{2k_{br}^2}\right]\exp\left[-\frac{Q^2}{2k_{br}^2}\right] \,,
\label{eq:Kqz}
\end{equation}
with 
\begin{align}
k_{br}^2&=\sqrt{\omega_0\hat q_0}, \hspace{0.2cm}&\omega_0=z(1-z)p_+, 
\hspace{0.2cm}&\hat q_0=\frac{\hat q}{N_c} f_{ji}(z)
\end{align}
and 
\begin{equation}
 f_{ji}(z)  =\frac{C_j + C_k - C_i}{2} + \frac{C_j + C_i - C_k}{2}\, (1-z)^2 + \frac{C_k + C_i - C_j}{2} \,z^2\,,   
\end{equation}
where the $C_i$ values are the squared Casimir operators of the color representations for the three correlated particles of the parton branching. 
Explicitly, we have
\begin{align}
f_{gg}(z)&=C_A\left[(1-z)+z^2\right]\,,\\
f_{qg}(z)&=C_F-C_Az(1-z)\,,\\
f_{gq}(z)&=C_A(1-z)+C_Fz^2\,,\\
f_{qq}(z)&=C_Az+C_F(1-z)^2\,.
\end{align}
A jet particle $i$ can also undergo scatterings off medium particles that transfer a momentum component $\mathbf{q}$ transverse to the incoming jet particle momentum at the time $t$. For these processes, we consider the following probability density
\begin{equation}
    \frac{\rmd^3 \mathcal{P}_{i}}{\rmd t\, \rmd^2 \mathbf{q}}=\frac{1}{(2\pi)^2}\,w_i(\mathbf{q})\,,
    \label{eq:scatrate}
\end{equation}
where the scattering kernel $w_i(\mathbf{q})$ in the case of gluons is given by
\begin{equation}
w_g(\mathbf{q})=\frac{16N_c\pi^2\alpha_s^2 n_{\rm med}}{\mathbf{q}^2(\mathbf{q}^2+m_D^2)}\,,
\label{eq:wg}
\end{equation}
with $n_{\rm med}$ being the density of scatterers and $m_D$ -- the Debye-mass. 
For jet quarks, the corresponding scattering kernel is obtained using the color factor $C_F/C_A$ as 
\begin{equation}
w_q({\bf q})=\frac{C_F}{C_A}w_g({\bf q})\,.
\label{eq:wq}
\end{equation}

In our current approach, all the individual jet particles undergo consecutive processes of medium-induced parton branching and scattering off medium particles at increasing times $t$. 
Once the time of a potential jet--medium interaction is larger than $t_L$, the time-scale that is necessary for the particles to pass a medium of size $L$, $t>t_L$, no further jet medium interactions are considered.
The jets after the in-medium evolution given by Eqs.~(\ref{eq:splitrate}) and ~(\ref{eq:scatrate}) can be obtained via the Monte Carlo algorithm described in more detail in Ref.~\cite{Rohrmoser:2021yrh}.
\section{Numerical results}\label{sec:numresults}
For our numerical results we simulate the productions of photon--jet pairs in ultrarelativistic heavy-ion collisions. 
To this end, we describe the production of photons and initial jet particles by the cross section for hard binary collisions given in Section~\ref{sec-ITMD}.
For this part, the \katie\ Monte Carlo program \cite{vanHameren:2016kkz} is used.
In order to study numerically the effects of the gluon saturation on observables of jets and photon--jet pairs, the hard binary collisions are obtained using the following different gluon TMDs which we describe below. For the more detailed discussion and initial conditions see the section nr. (6) of the recent review \cite{vanHameren:2023oiq}:
\begin{itemize}
    \item TMD for Pb that takes into account the  saturation effects in the lead core according to Ref.~\cite{Kutak:2012rf}.
    This gluon density is a solution of the Balitsky--Kovchegov (BK) \cite{Balitsky:1995ub,Kovchegov:1999yj} equation with supplementary corrections coming from the kinematical constraint and the non-singular pieces of the DGLAP splitting function. This terms allow for resummation of subleading in $\ln 1/x$ effects and make the equation phenomenologically applicable. 
    It takes into account also the Sudakov effects 
     \cite{Al-Mashad:2022zbq} that allow for resummation of effects at relatively large $x$ and small scales.
    The nonlinear term introducing gluon recombination is multipled by parameter encoding information of size of the target and atomic number, in this case A=208.
    
    \item TMD for the proton that takes into account the saturation ~\cite{Kutak:2012rf}.  While the evolution and the initial condition is exactly the same as in the one above, it differs form it by strength of a nonlinear term determined by the size of the target, in this case $A=1$. We refer to it as {\tt p KS}. We need it to calculate nuclear modification ratio. 
      \item TMD that does not account for the effects of the gluon saturation from Ref.~\cite{Kutak:2012rf}. This gluon density is a solution of the Balitsky--Fadin--Kuraev--Lipatov (BFKL) equation and similarly as the TMD above it accounts for  corrections coming from a kinematical constraint and non-singular pieces of the DGLAP splitting function \cite{Kwiecinski:1997ee}. We refer to this TMD as {\tt KSlin}.

    \end{itemize}
    
For the collinear and large $x$ parton in the hard binary collisions, we use the nCTEQ15 nuclear PDFs for $^{208}$Pb from Ref.~\cite{Kovarik:2015cma}. 
The in-medium jet propagation is obtained by including the processes of VLEs as well as scatterings and induced radiation, as described in Subsections~\ref{sec:vle} and~\ref{sec:medium}. This part is achieved by using the partons produced in the hard collision events by \katie~ as input for the \tmdice~ algorithm that allows to obtain jet particles after the in-medium evolution.
 We obtain the numerical results for the in-medium jet-evolution with the set of parameters given in Tables~\ref{tab:simpar} and~\ref{tab:medpar}. 
 For the description of jet--medium interactions a medium of length $t_L$ with constant temperature $T$ was presumed and the values for $n_{\rm med}$, $m_D$, and $\hat{q}$ that are necessary for the splitting and scattering kernels in Eqs.~(\ref{eq:Kqz}) and~(\ref{eq:wg}), respectively, are obtained with the same parameterization as previously in Ref.~\cite{vanHameren:2019xtr}. Table~\ref{tab:medpar} gives the corresponding results for $T=250\,$MeV.
 
 \begin{table}[ht]
     \centering
     \begin{tabular}{||c|c|c|c|c|c|c||}
     \hline\hline
                 $t_L$  & $\bar{\alpha}=\frac{\alpha_s}{\pi}$ & $x_{\rm min}$ & $k_{\perp\,\,{\rm min}}$&  $x_{\rm min\, VLE}$ &$\epsilon_{\rm VLE}$ & $R_{\rm max}$  \\\hline
      $5\,$fm/c & $0.1$ & $10^{-3}$  & $0.5\,$GeV & $10^{-2}$&$10^{-4}$ & $1$ \\
      \hline\hline
     \end{tabular}
     \caption{Numerical parameters for the jet-evolution via \tmdice.}
     \label{tab:simpar}
 \end{table}

  \begin{table}[ht]
     \centering
     \begin{tabular}{||c|c|c||}
     \hline\hline
                $n_{\rm med}$ & $\hat{q}$& $m_D$  \\\hline
      $0.08\,$GeV$^3$ & $0.29\,$GeV$^2/$fm& $0.61\,$GeV\\
      \hline\hline
     \end{tabular}
     \caption{Parameters for the description of a uniform medium with the temperature $T=250\,$MeV.}
     \label{tab:medpar}
 \end{table}
In order to be able to construct the jet and jet--photon observables from our data for PbPb collisions we obtain jets and jet-momenta from the simulated numerical data by means of the anti-$k_T$ algorithm (with the parameter $R=0.2$) in the implementation provided by the {\sf FASTJET} program~\cite{Cacciari:2011ma,Cacciari:2005hq}.

\begin{figure}[ht]
    \centering
    \includegraphics[width=0.50\textwidth]{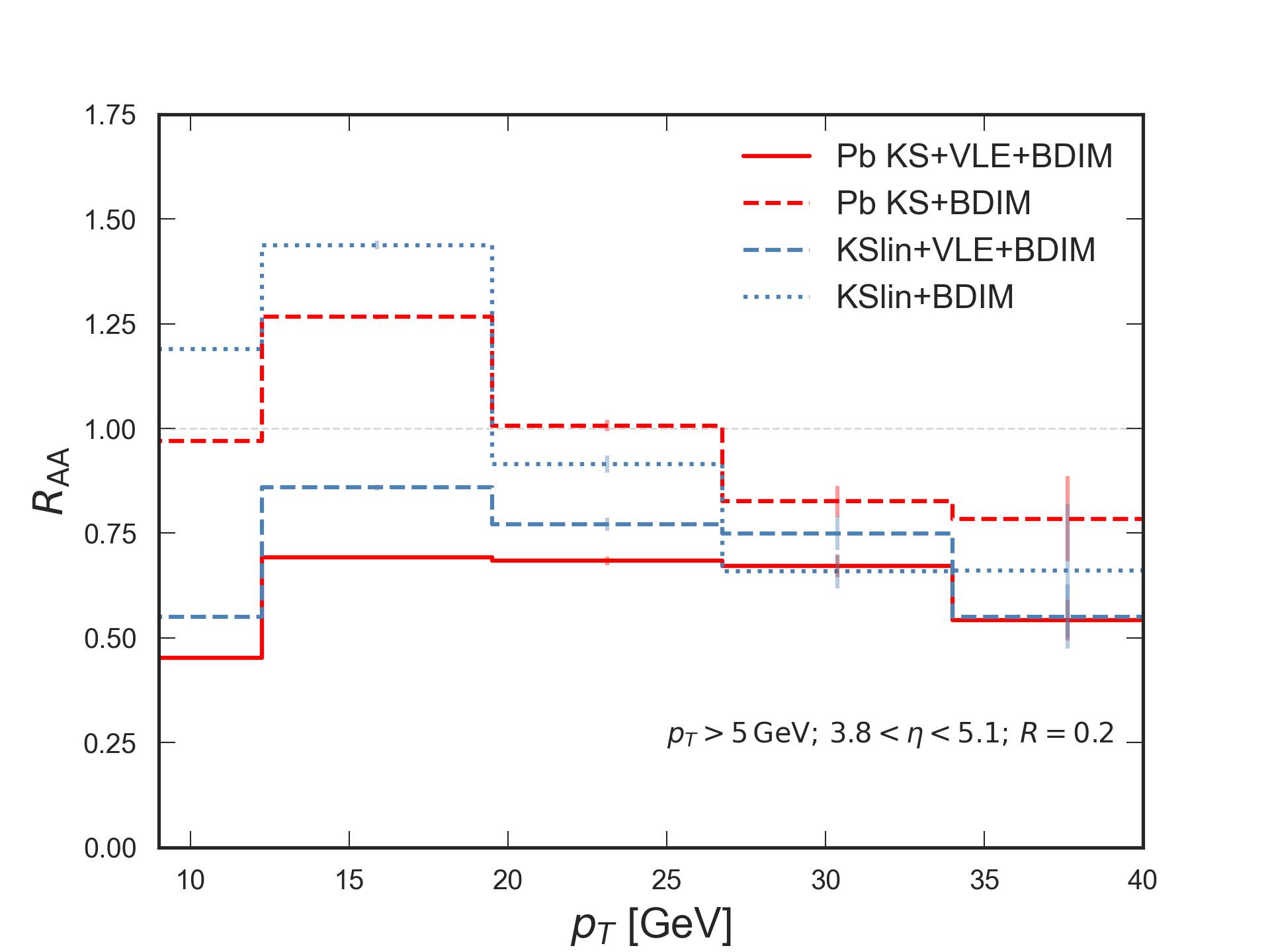}
    \caption{The nuclear modification factor $R_{AA}$ for anti-$k_T$ jets with $R=0.2$ produced in PbPb collisions after the in-medium evolution via VLEs, scatterings and medium-induced radiations. The different results correspond to $p$ and Pb TMDs with and without the saturation and the Sudakov resummation, for the following acceptances: $3.8 < y < 5.2$, $p_{T}^{ini,\gamma} \ge 5\,$GeV and $p_{T} \ge 5\,$GeV.}
 \label{fig:jet-RAA_vle_bdim} 
\end{figure}
A widely studied inclusive observable is the jet nuclear modification factor $R_{AA}$ defined as 
\begin{equation}
    R_{AA}(p_T)=\frac{dN_{AA}/dp_T}{\langle N_{coll}\rangle\, dN_{pp}/dp_T}\,,
    \label{eq:RAA1}
\end{equation}
where $N_{AA}$ and $N_{pp}$ are the number of jets produced in AA and $pp$ collisions, respectively, and $\langle N_{coll}\rangle$ is the average number of binary nuclear collisions.
For the qualitative considerations in this paper, nuclear effects of enhancement or suppression in jet-production other than the saturation have been neglected.
To be precise to get the numerator of $R_{AA}$ for $\sigma_{AA}$ we employ for the low $x$ TMD the KSlin and PBKS while for the denominator the pKS. 

Given above, we approximate the jet nuclear modification factor as
\begin{equation}
    R_{AA}(p_T)\approx\frac{d\sigma_{AA}/dp_T}{d\sigma_{pp}/dp_T}\,,
    \label{eq:RAA2}
\end{equation}
with $\sigma_{AA}$ and $\sigma_{pp}$ being the cross sections for jet production in AA and $pp$ collisions, respectively. 
For the $pp$ collisions, we also estimate the effects of jet-energy loss in the vacuum by QCD emissions via small-$R$ resummation~\cite{Dasgupta:2016bnd}. Numerically, this is achieved by first obtaining quark--photon pairs from Monte Carlo simulations of $pp$ collisions, similarly as in Ref.~\cite{Ganguli:2023joy}. For this, we use the \katie\ program. Then, the produced quarks are used as initial particles of parton cascades with several consecutive emissions from the quarks that follow the DGLAP evolution equations. This step is done using the \tmdice\ program. The branching angles lie inside the range between $R=0.2$ and $R_{\rm max}=1$. Whenever the emission at the branching angle lower than $R$ is selected by the Monte Carlo algorithm, the evolution is stopped and the four-momentum of the decaying quark is considered as an estimate of the four-momentum of a jet after radiation of a parton outside a cone of the size $R$.

Figure~\ref{fig:jet-RAA_vle_bdim} shows the jet nuclear modification factor of forward jets in the region $3.8<\eta<5.1$ for different gluon TMDs with and without the saturation.
As can be observed, the results that account for VLEs are suppressed at low values of $p_T$ compared to those that neglect them.
Furthermore, at low enough values of $p_T$ (below $p_T=20\,$GeV, when VLEs are not included, and below $p_T=35$~GeV, when they are included), a stronger suppression can be observed for the cases with the saturation. At higher values of $p_T$, one observes, at least for the scenario where VLEs are included, that the case with the saturation shows a smaller suppression than the case without it. 

The inclusion of VLEs strongly increases the jet-suppression at the low-$p_T$ values by an approximate factor of $2$ for $p_T\le 20$~GeV. 
The relatively strong effect of VLEs at these $\pT$ values can be attributed to the fact that the jet \textit{energy} is quite high, $E \gg \pT$, opening up the phase space for multiple emissions.

\begin{figure}[t!]
    \centering
    \includegraphics[width=0.5\textwidth]{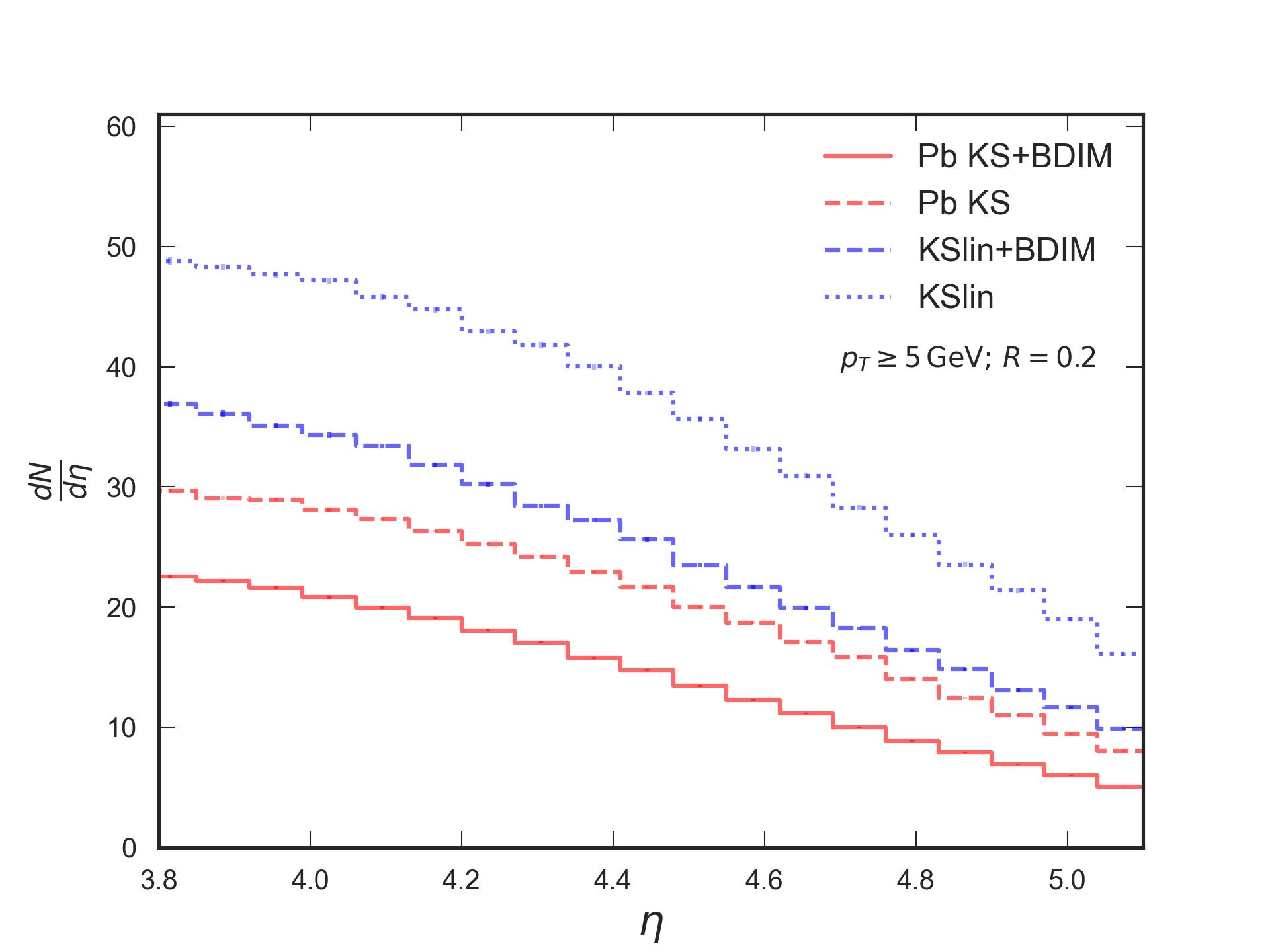}\hspace{-0.5cm}
    \includegraphics[width=0.5\textwidth]{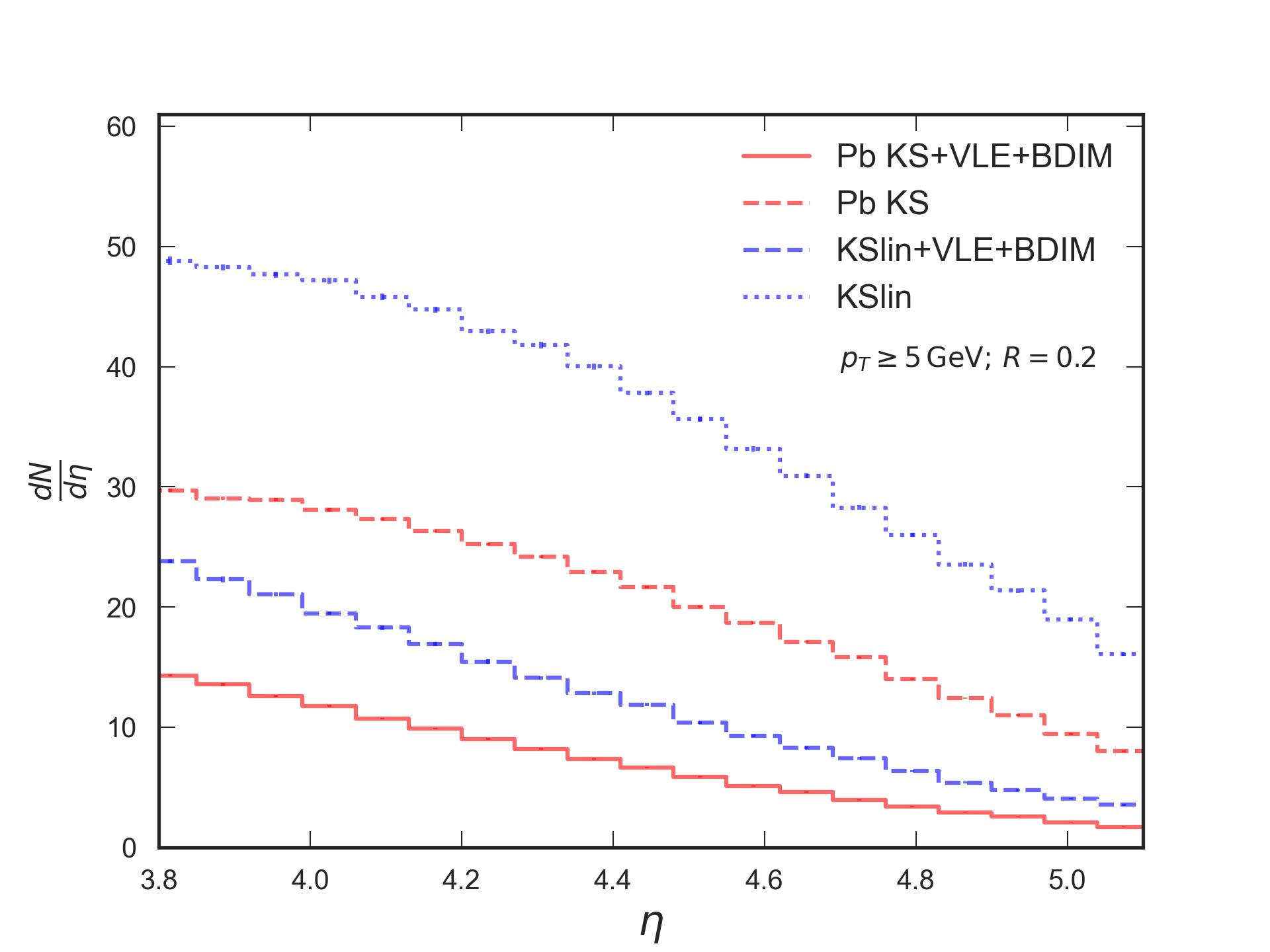}
    \caption{Comparisons of the rapidity distribution of jets for two different configurations of no-medium and medium case for the $\gamma$-jet input from PbPb collisions at $\sqrt{s} = 5.02\,$TeV.
    The TMD grids are for all possible combinations of the saturation and Sudakov effects. The acceptances are: $3.8 < y < 5.1$, $p_{T}^{ini,\gamma} \ge 5\,$GeV and $p_{T}\ge 5\,$ GeV.} 
    \label{fig:deta}\label{fig:detacomp}
\end{figure}

Figure~\ref{fig:deta} compares the distributions in rapidities of jets and photons obtained for the cases of in-medium jet propagation to the cases without in-medium jet-propagation obtained directly from \katie\ for the different gluon TMDs. 
The suppression due to non-linearities is visible in all the considered scenarios. The nonlinear effects slow down the energy evolution of gluon density which is than visible in lower slope of rapidity distribution of results with saturation.
Furthermore as can be seen, the presence of the jet--medium interactions leads overall to a considerable suppression of the rapidity distributions in comparison to the cases without the in-medium evolution.
While the upper panel of Fig.~\ref{fig:deta} shows the rapidity distributions for jet particles undergoing only in-medium scatterings and medium-induced radiations, in the lower panel, the distributions that additionally include VLEs are presented. When the corresponding cases of the jet--medium interactions with and without VLEs are compared, the distributions with VLEs exhibit a stronger suppression for all the considered TMDs. Overall, the scenario that includes both the saturation and VLEs gives the largest suppression. 
\begin{figure*}[ht!]
    \centering
    \includegraphics[width=0.5\textwidth]{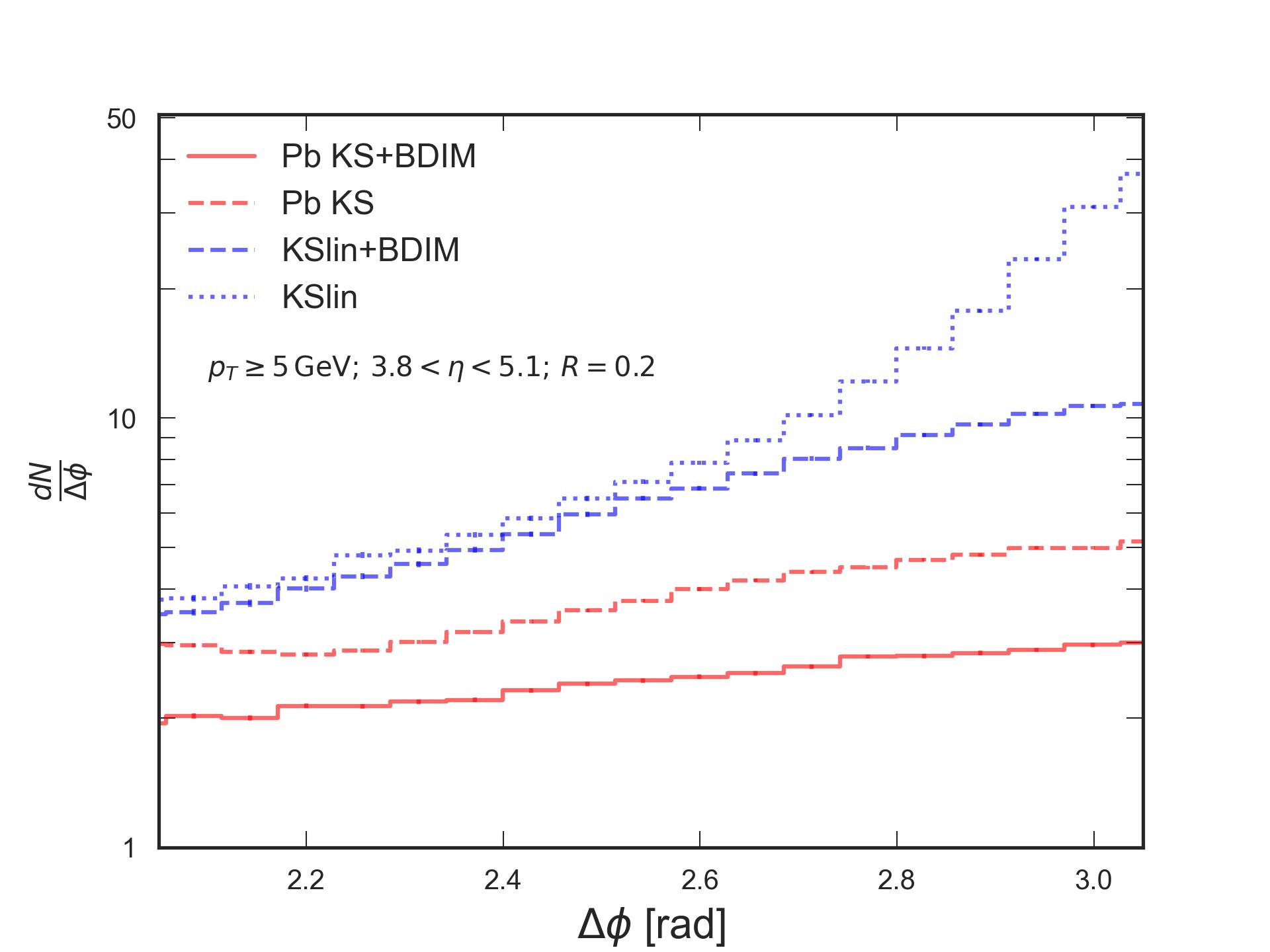}\hspace{-0.5cm}
    \includegraphics[width=0.5\textwidth]{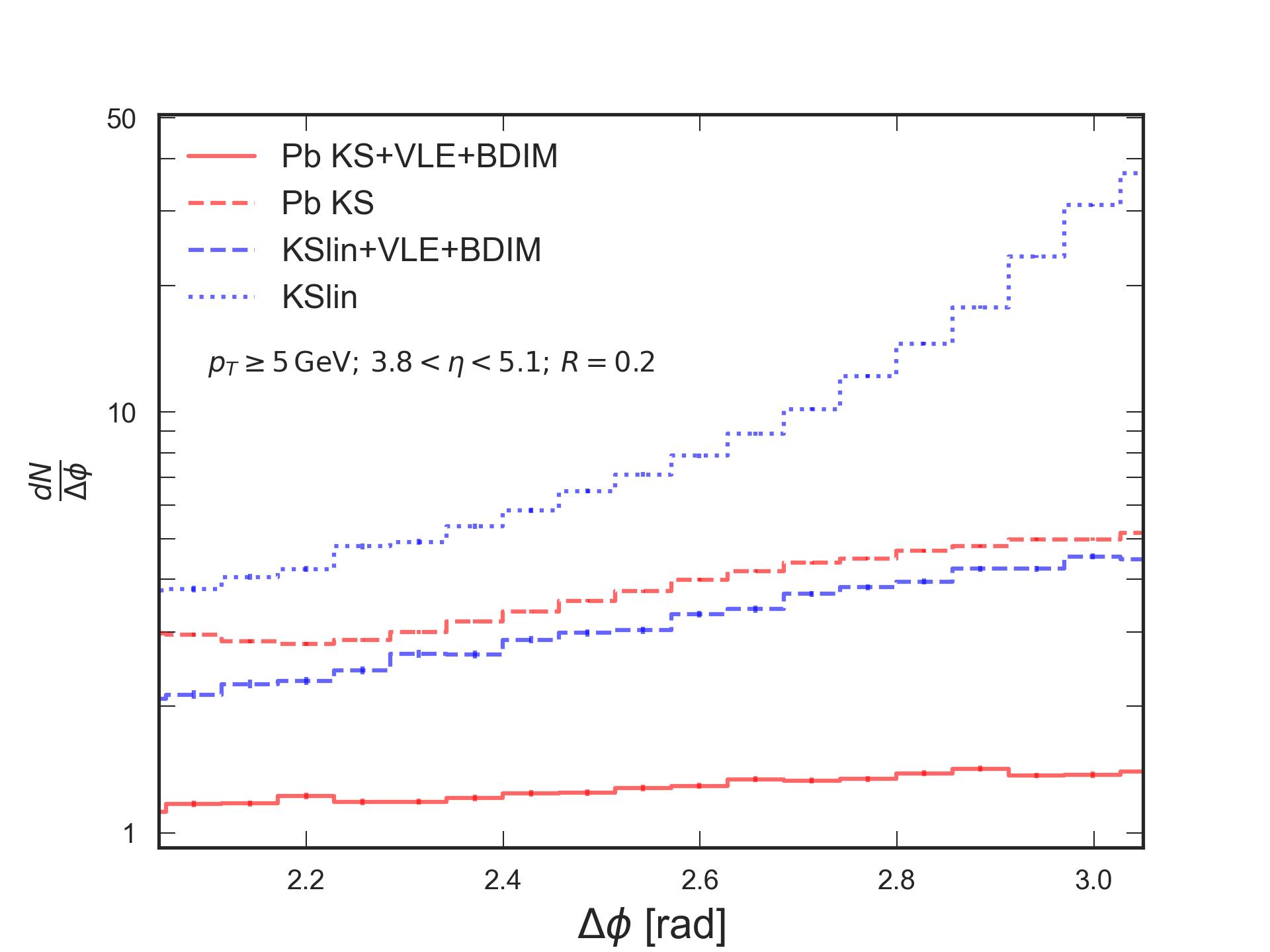}\\
    \includegraphics[width=0.5\textwidth]{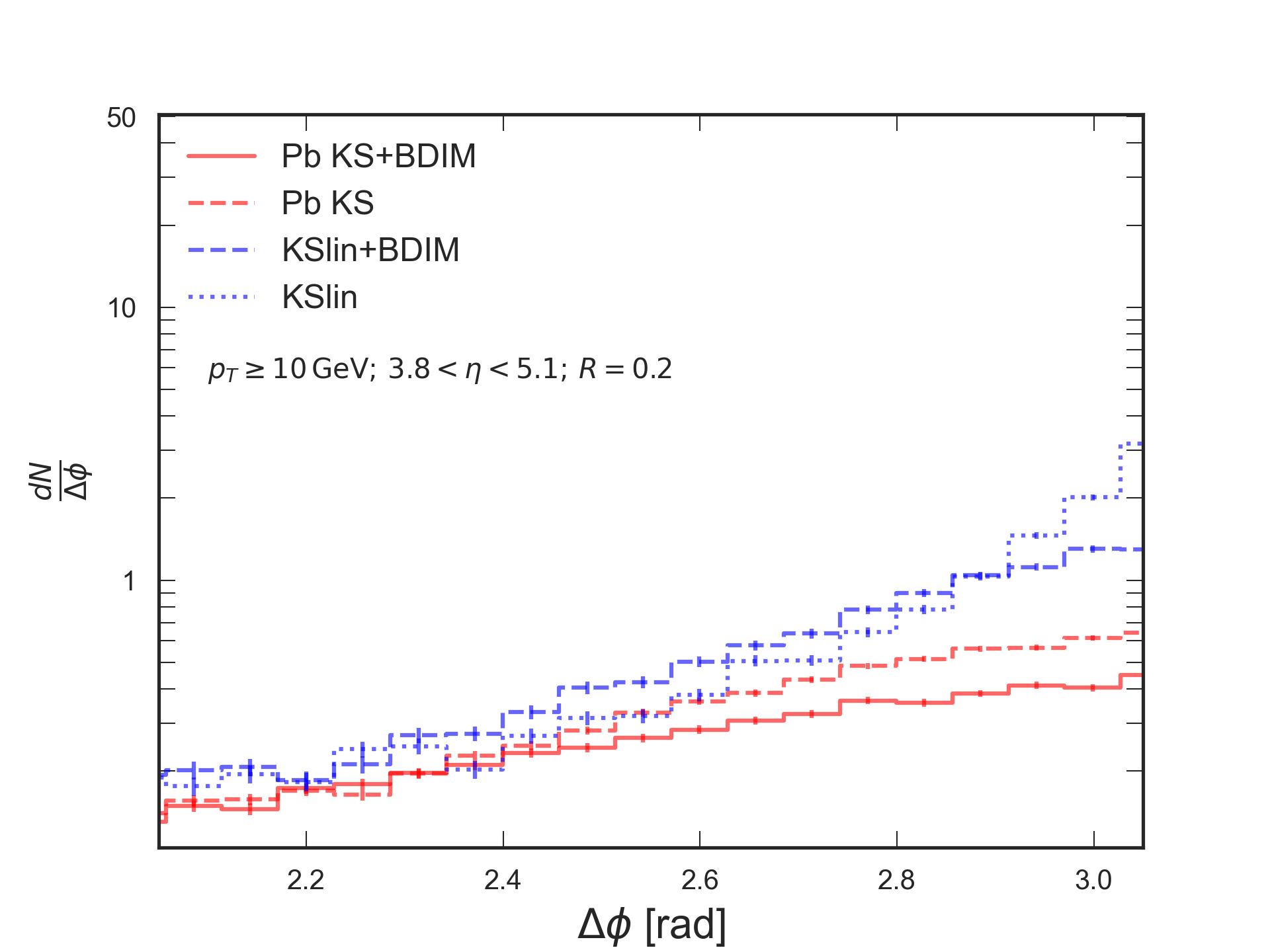}\hspace{-0.5cm}
    \includegraphics[width=0.5\textwidth]{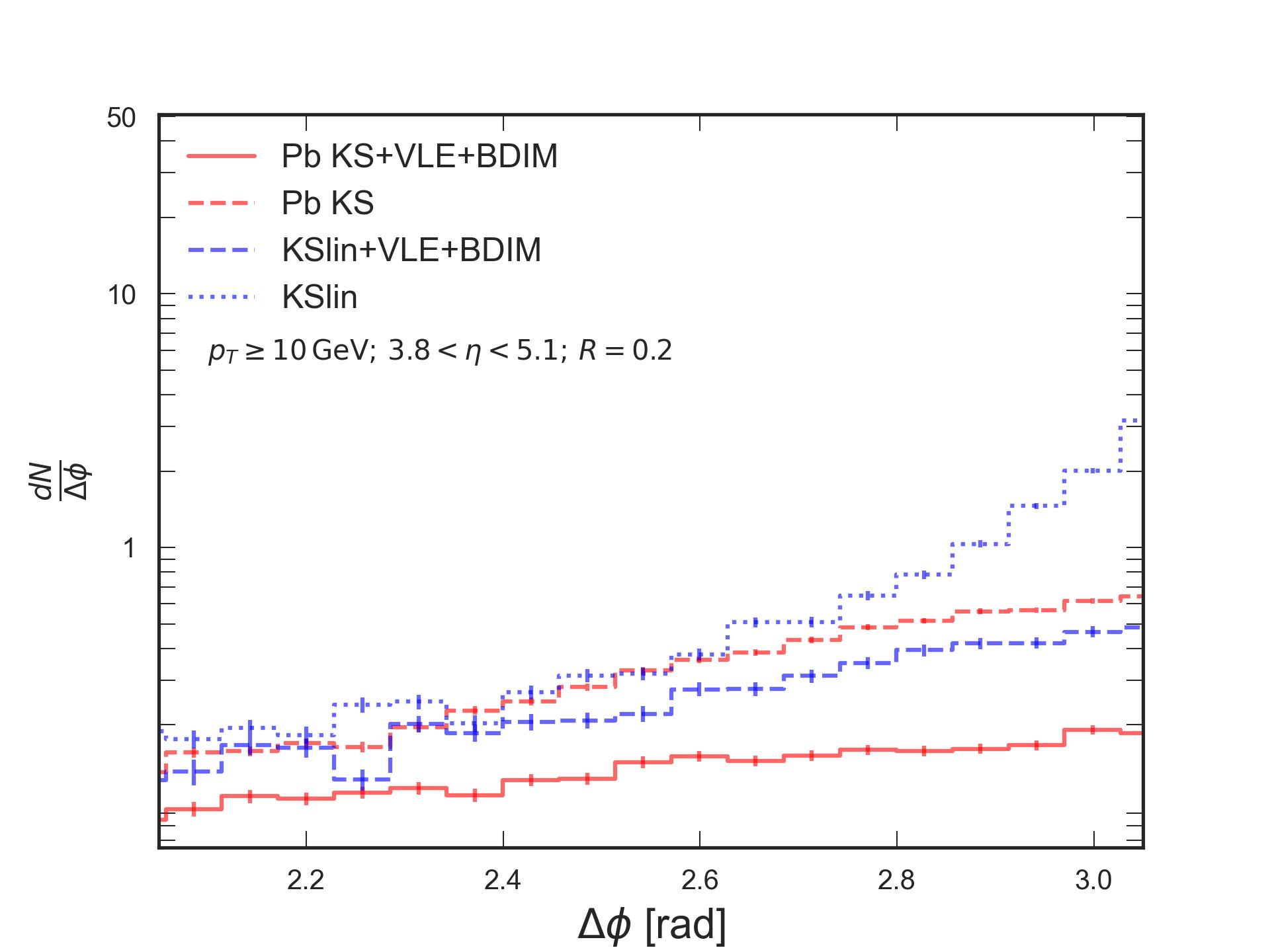}
        \caption{Comparisons of the azimuthal-angle decorrelations between jets and photons for different configurations of no-medium  and medium case for $\gamma$--jet input from PbPb collisions at $\sqrt{s} = 5.02\,$TeV. 
        The TMD grids are for all possible combinations of the saturation and Sudakov effects. The acceptances are: $3.8 < y < 5.2$, $p_{T}^{ini,\gamma} \ge 5\,$GeV with $p_{T} \ge 5\,$GeV (top panel) and $p_{T} \ge 10\,$GeV (bottom panel).  
        }
    \label{fig:dphibdim}\label{fig:dphivlebdim}
\end{figure*}

Figure~\ref{fig:dphibdim} shows the azimuthal-angle decorrelations for the jet--photon pairs, where the jets follow the BDIM evolution,
in comparison to results without the in-medium jet evolution. 
The cases with the jet evolution exhibit a strong suppression at $\Delta\phi=\pi$, which becomes significantly less pronounced at smaller values of $\Delta\phi$ where the distributions with and without the jet--medium interactions start to converge. 

First of all  one can observe much more suppressed distributions for the TMDs with the saturation  than for the results with KS-linear TMD. This difference occurs both for the cases with and without the in-medium jet-evolution. 
We attribute this effect to saturation effects  which are consequence of gluon recombination at low $x$.
In the nearly back to back region the imbalance, i.e.\ $k_T$ of the off-shell initial-state gluon is small. In the saturation scenario such low $k_T<Q_s$, where $Q_s$ is a saturation scale,  gluons are suppressed leading to suppression of the cross-section. This is however not the case in the linear case.  Therefore we see such pronounced difference between the considered scenarios. Now as one moves toward the smaller $\Delta\phi$ one is probing gluons with larger $k_T$ and the differences diminish.

Furthermore the suppression comes from medium jet interaction. The invariance under the jet--medium interactions appears at larger values of $\Delta\phi$, if one rejects all the jet--photon pairs where $p_T$ of either the jet or the photon is below a certain value. 
One can explain such a behavior by the suppressed jet--medium interactions for jet-particles at large $p_T$. The observed behavior can be expected from a BDMPS-Z-type of coherent medium-induced emissions which are suppressed by a factor of $1/\sqrt{\omega}$ (where $\omega$ is the energy of the emitter).

The results for the two observables discussed above, $\eta$ and $\Delta\phi$, indicate that both spectra are informative and carry valuable insights. Overall, our results show that saturation effects are visible even if the jet interact with medium. This suggest that the saturation effects may be observable in experimental analyses of jet-quenching in heavy-ion collisions.
\section{Summary and conclusions}
\label{sec:conclusions}

In summary, this work presents the first study of jet-quenching in the forward acceptance region of collider detectors, utilizing a combination of gluon-saturation dynamics and in-medium jet evolution. 
We find that the combination of non-perturbative saturation phenomena at early-stages followed by a perturbative regime through jet fragmentation may have a chance to be  visible in spectra of the studied final states.
As a concrete example, we have provided  predictions suited to the kinematical cuts of the planned forward-calorimeter detector experiments. Furthermore, we have included a detailed description of the in-medium jet propagation through VLEs followed by the in-medium parton-branching process leading to the jet energy loss and intra-jet modifications. By comparing scenarios that combine the jet-quenching with the gluon saturation to those with the quenching but without the saturation, we study the impact of saturation on final-state observables. While the measurement of the considered final state is more challenging than the $p$Pb case, where the cross section does not get suppression from the quark-gluon plasma \cite{Ganguli:2023joy}, it is complementary to the latter and will be measured in the forthcoming upgrades of the ALICE detector.
Furthermore, the considered process is in a way simpler than the central collision as one can use the hybrid kinematics  that simplifies the factorization.
 Finally, we demonstrate that VLEs provide the additional suppression of the cross section in nuclear collisions.

In addition to the above, a qualitative understanding of the jet production in the $\gamma$--jet final state in the forward region provides an effective tool to study the flavor dependence in the forward rapidities, it also allows for an in-depth study of the role of selection biases as well as the path length dependence of the in-medium jet energy loss \cite{Zhang:2009rn,Qin:2009bk,Renk:2006qg}, as also pursued extensively in~\cite{Betz:2014cza,Betz:2016ayq,Djordjevic:2018ita,Arleo:2022shs}.

The detailed methodology to include  the saturation effects and the in-medium quenching presented here for the $\gamma$--jet configuration can be readily extended for generic dijet systems. In our study we have neglected glasma fields \cite{Lappi:2006fp} which are expected to be relevant for early stages of heavy ion collisions and can lead to suppression of angular correlations \cite{Avramescu:2024xts}. The are however mostly contributing for processes where both of colliding ions are saturated and final states are produced in central rapidity region \cite{McLerran:2008es,Carrington:2021dvw,Carrington:2022bnv}.

As an outlook, our study can be further developed by considering the possibility for implementing splitting kernels that account for a more realistic expansion scenario \cite{Adhya:2019qse}, as well as study through the framework of the resummed opacity expansion approach \cite{Isaksen:2022pkj}. One can also study the same set up with recent Monte Carlo frameworks for modeling the jet--medium interactions, such as {\sf SUBA-jet} \cite{Karpenko:2024fgg}.
Additionally, one could explore further by including the energy-smearing effects from the initial state radiation, as in Refs.~\cite{Wang:1996yh,Wang:1996pe}, for which the jet spectra can be significantly modified leading to a sizable effect on the jet nuclear modification factor $\raa$. In this study, we have restricted ourselves to the lowest order in $\alpha_s$, leading to the transverse energy smearing function as a $\delta$-function due to the momentum conservation. 

To further advance our study, one could include the other channels for $\gamma$--jet production, such as $gq^*\rightarrow \gamma q$ and $q\bar{q}^*\rightarrow \gamma q$, in addition to the $qg^*\rightarrow q\gamma$ that we have considered here. However, they are subleading and should not influence much the main results of this paper. We shall incorporate them together with other higher-order effects in our more detailed study in the future.

\section*{Acknowledgements}

 SPA acknowledges funding from grant agreement PAN.BFD.S.BDN.612.022.2021-PASIFIC 1, QGPAnatomy and Physics for Future (P4F) Project No. 101081515, PEAKIN. This work received funding from the European Union’s Horizon 2020 research and innovation program under the Maria Sklodowska-Curie grant agreement No.\ 847639 and the Polish Ministry of Education and Science. This work received funding from the Horizon Europe Programme, HORIZON-MSCA-2021-COFUND-01 call by the European Research Executive Agency.
KT acknowledges the partial support by
(NCN) grant 2023/50/A/ST2/00224.
This work received funding from the European Union’s Horizon 2020 research and innovation programme under the Maria Sk{\l}odowska-Curie grant agreement No.\ 847639 and from the Polish Ministry of Education and Science. 
The research of MR has been supported by the Polish National Science Centre (NCN) grant no.\ DEC-2021/05/X/ST2/01340. 
The research of WP has been supported in part by a grant from the Priority Research Area (DigiWorld) under the Strategic Programme Excellence Initiative at the Jagiellonian University in Krakow, Poland.

\appendix
\section{Uncertainties in observables}
In order to estimate the influence of the jet-medium interactions on the studied observables, we have obtained some sample results for a medium temperature of $T=187.5$~MeV. 
The resulting values for the parameters $\hat{q}$, $n_{\rm med}$ and $m_D$ are:

     \begin{tabular}{||c|c|c||}
     \hline\hline
     \centering
                $n_{\rm med}$ & $\hat{q}$& $m_D$  \\
                \hline
      $0.0338\,$GeV$^3$ & $0.1223\,$GeV$^2/$fm& $0.4575\,$GeV\\
      \hline\hline
     \end{tabular}\\

\begin{figure}[ht]
    \includegraphics[width=0.5\textwidth]{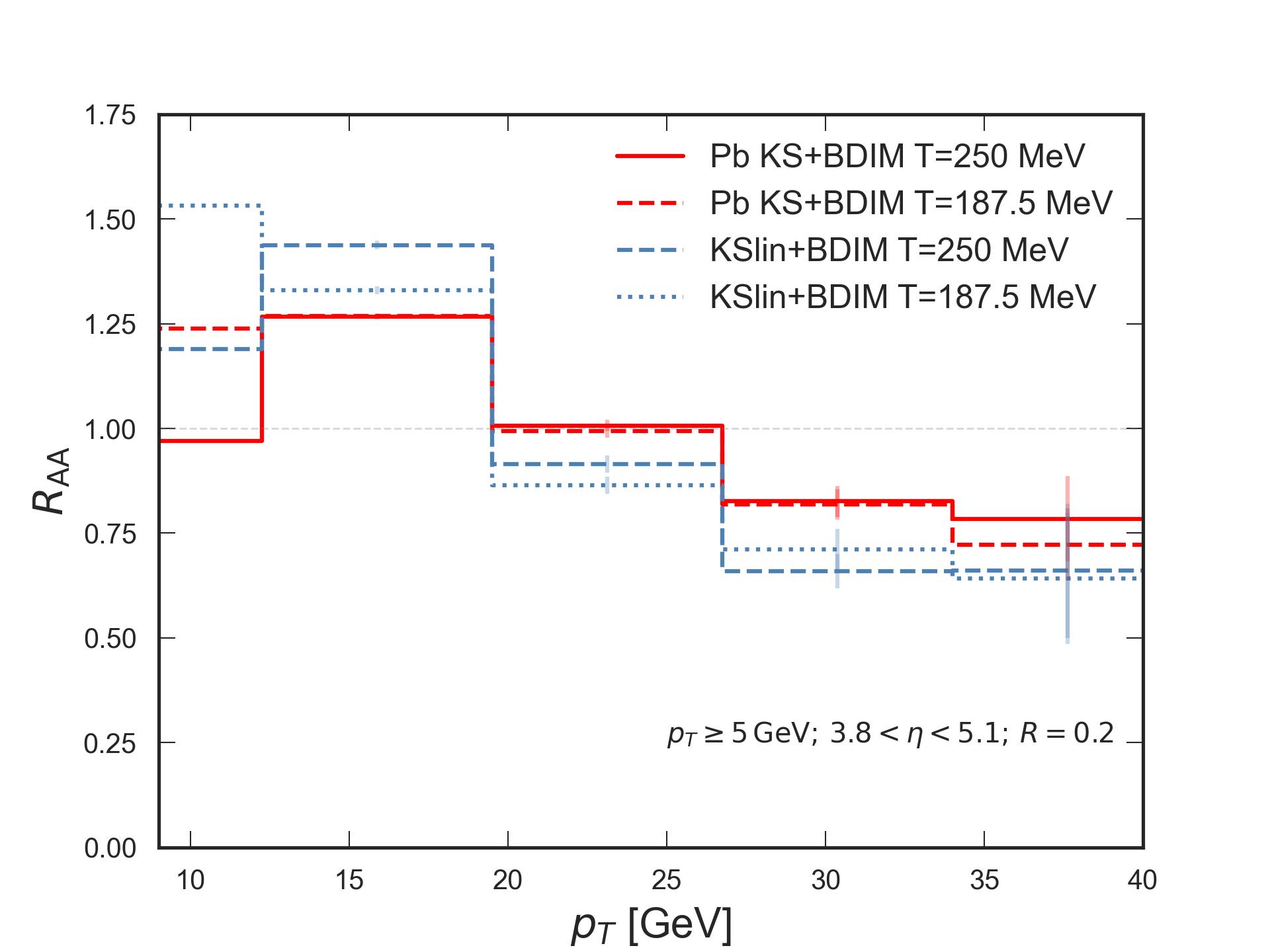} 
    \includegraphics[width=0.5\textwidth]{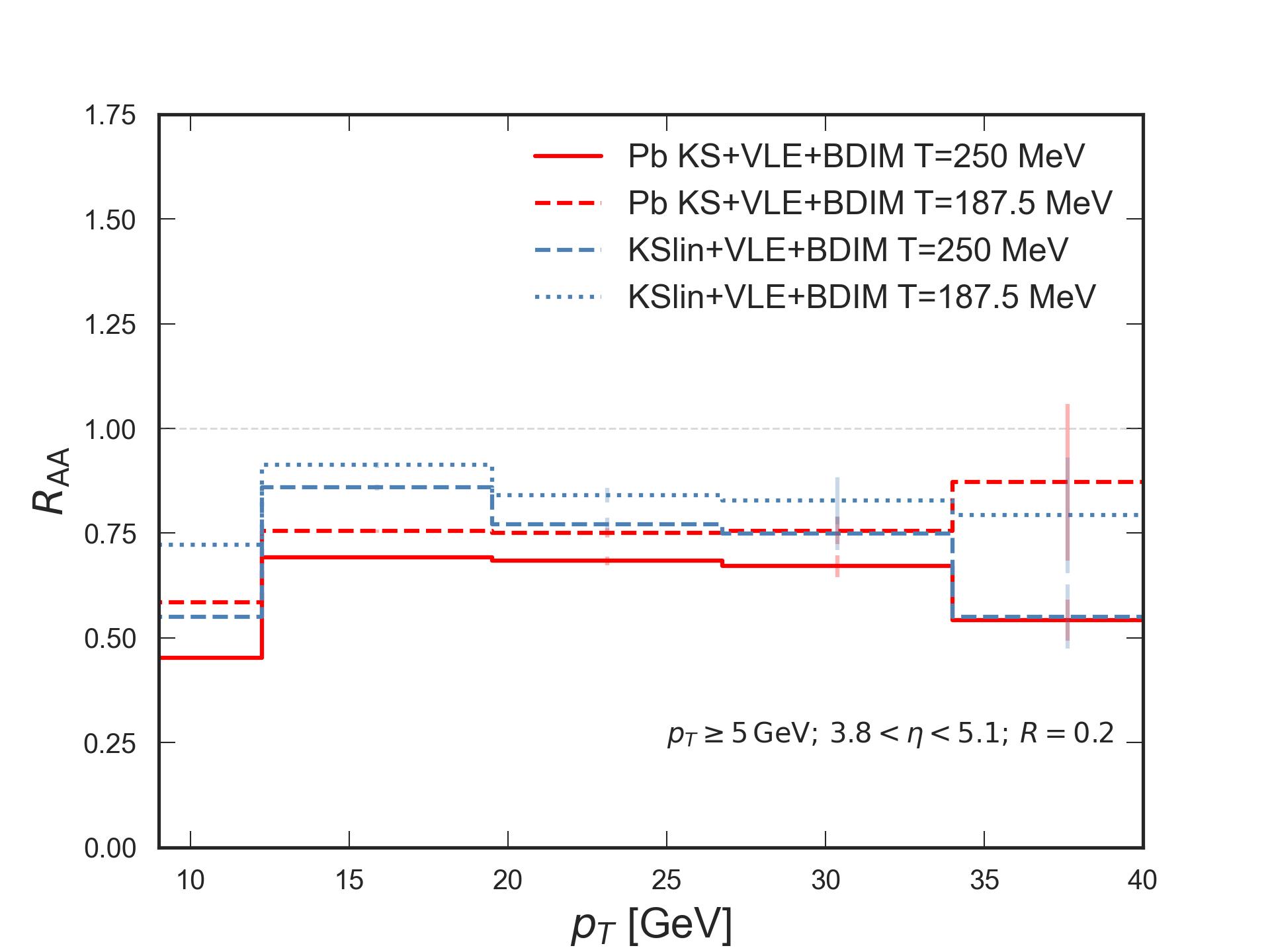}
    \caption{Nuclear modification factors for media with the temperatures of $T=187.5\,$MeV and $T=250\,$MeV, respectively.}
    \label{fig:raa_unc}
\end{figure}

\begin{figure}[ht]
    \includegraphics[width=0.5\textwidth]{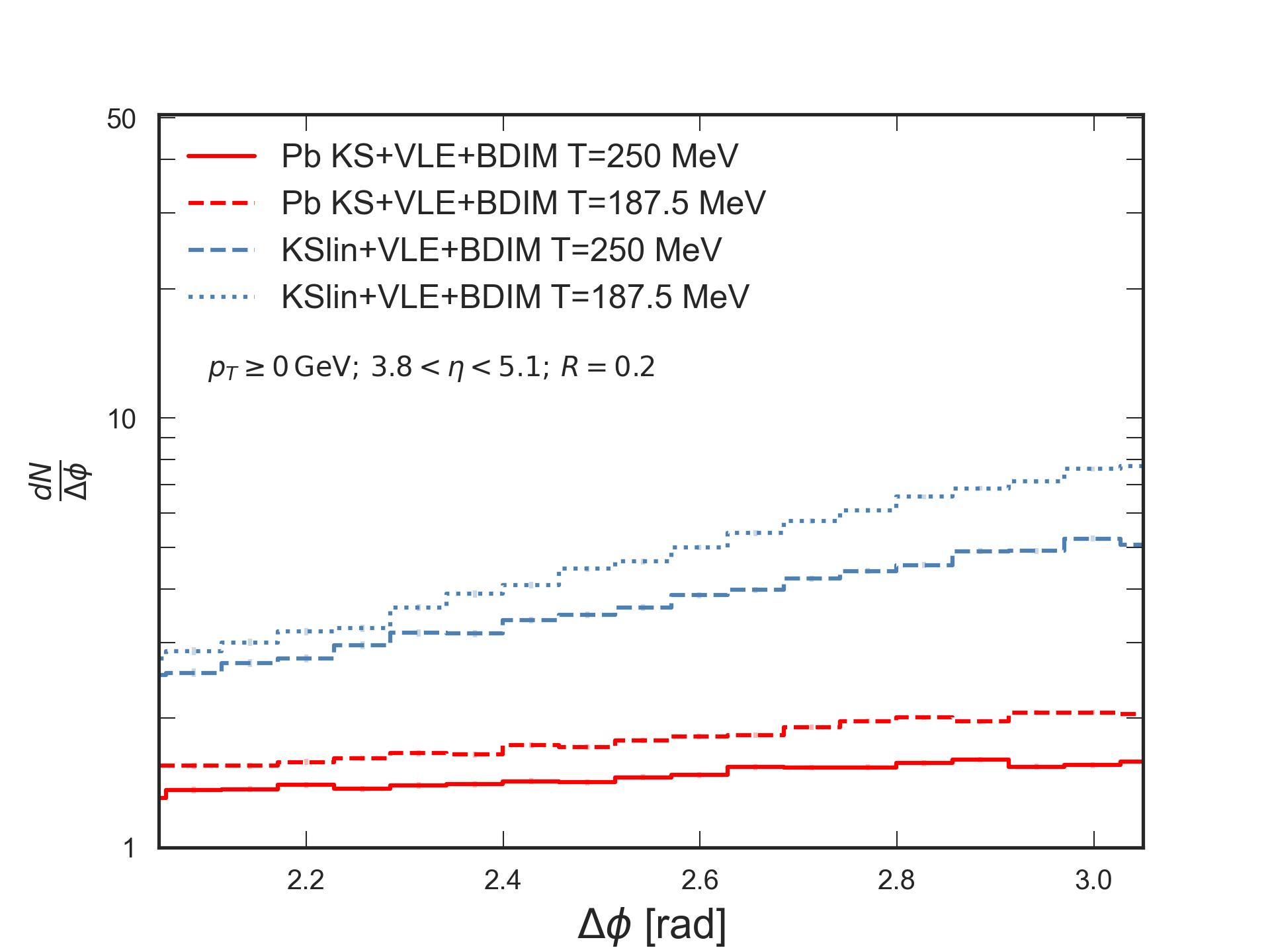} 
    \includegraphics[width=0.5\textwidth]{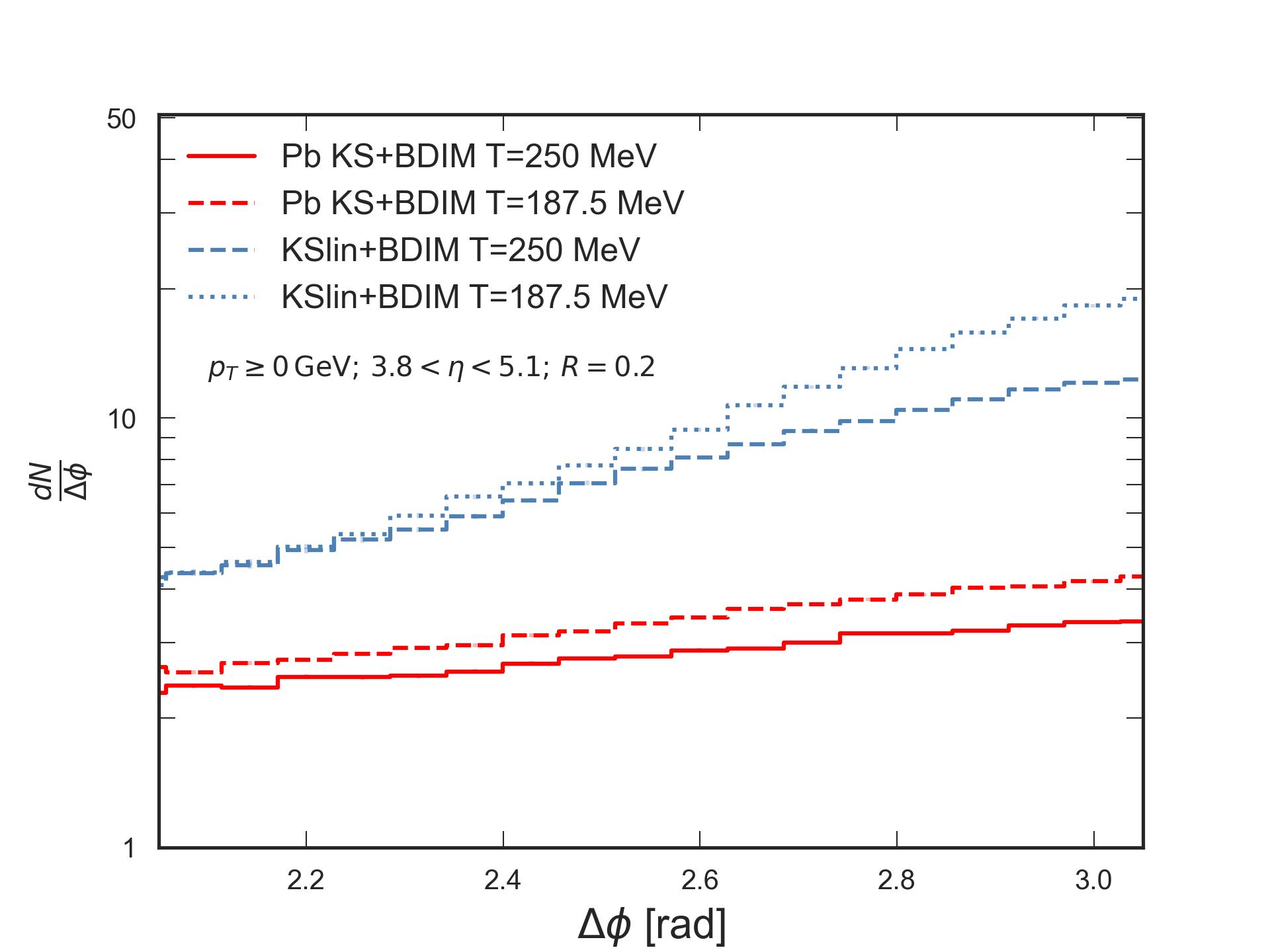}
    \caption{Azimuthal decorrelations for media with the temperatures of $T=187.5\,$MeV and $T=250\,$MeV, respectively.}
    \label{fig:dphi_unc}
\end{figure}

Fig.~\ref{fig:raa_unc} shows the results for the jet nuclear modification factor $R_{AA}$ for jet evolution with VLEs and without VLEs in comparison to the previously shown results for $T=250$~MeV. As can be seen, the differences between the values of $\raa$ with and without inclusion of the gluon saturation are in general larger than the differences resulting from a $25$\% change of the medium temperature. Similar observations can be seen in the comparisons shown in Fig.~\ref{fig:dphi_unc} for the azimuthal-angle decorrelations.

\section{Monte Carlo algorithm for VLEs}
\label{App:VLE}

In \tmdice, VLEs evolve following the DGLAP evolution equations for time-like final-state cascades ordered in virtuality $Q^2$, i.e.
\begin{align}
    Q^2\frac{\partial}{\partial Q^2}D_q(x,Q^2)&= \frac{\alpha_s}{2\pi}\int dz P_{qq}(z)\left[\frac{1}{z}D_q\left(\frac{x}{z},Q^2\right)-D_q(x,Q^2)\right]\nonumber\\
    &+\frac{1}{z}P_{qg}(z)D_g\left(\frac{x}{z},Q^2\right)\,,
    \label{eq:DGLAPq}\\
    Q^2\frac{\partial}{\partial Q^2}D_g(x,Q^2)&= \frac{\alpha_s}{2\pi}\int dz P_{gg}(z) \frac{1}{z}D_g\left(\frac{x}{z},Q^2\right)
    \nonumber\\&+ P_{gq}(z) \frac{1}{z}D_q\left(\frac{x}{z},Q^2\right)-\left[P_{gg}(z)+N_F P_{qg}(z)\right]\nonumber\\&\times D_g(x,Q^2)\,,
    \label{eq:DGLAPG}
\end{align}
with the gluon and quark fragmentation functions $D_g(x,Q^2)$ and $D_q(x,Q^2)$, respectively.
As with the BDIM evolution of jets, the running of the QCD coupling was neglected and $\alpha_s$ is given as a constant input parameter.
The DGLAP evolution equations can be solved numerically via a Monte Carlo algorithm. 
To this end the Sudakov functions for the quark and the gluon are defined, respectively as 
\begin{align}
\Delta_q(Q_1^2,Q_2^2,E)&=\exp\left[-\int_{Q_2^2}^{Q_1^2}\frac{dQ^2}{Q^2}\int_{x_-(Q^2,E,R_{\rm max})}^{x_+(Q^2,E)} dz \frac{\alpha_s}{2\pi}P_{qq}(z)\right]\,,\label{eq:Sudq}\\
\Delta_g(Q_1^2,Q_2^2,E)&=\exp\bigg[-\int_{Q_2^2}^{Q_1^2}\frac{dQ^2}{Q^2}\int_{x_-(Q^2,E,R_{\rm max})}^{x_+(Q^2,E,R_{\rm max})} dz  \frac{\alpha_s}{2\pi} P_{gg}(z)
\nonumber\\&+N_F P_{qg}(z)\bigg]\,,\label{eq:SudG}
\end{align}
{where $E$ is the energy of the quark or gluon, and we have introduced the integration boundaries $x_\pm(Q^2,E,R_{\rm max})$, such that all possible parton splittings into particles on and off the mass-shell at angles smaller than a maximal angle $R_{\rm max}$ are considered. These boundaries are then given as
\begin{equation}
    x_\pm(Q^2,E,R_{\rm max})=\frac{1}{2}\left[1\pm\sqrt{1-\frac{2Q^2}{E^2(1-\cos (R_{\rm max}))}}\,\right]\,.
\end{equation}

For partonic cascades, the above integration boundaries would allow to consider angular ordering, if $R_{\rm max}$ is always adapted to the previous branching angle. In the current work, however, $R_{\rm max}$ is considered as a global maximal angle given as a parameter in Table~\ref{tab:simpar}.
Differences between this procedure and strict angular ordering could lead to discrepancies at higher logarithmic order, but we have not quantified this further with the current version of \tmdice.

The following CDFs are defined for the splitting functions of a particle $j$ into a particle $i$
\begin{equation}
    \rho_{ij}(z,Q_i^2,E_i)=\int_{x_-(Q_i^2,E_i,R_{\rm max})}^zd\tilde{z}P_{ij}(\tilde{z})\,.
\end{equation}

It is then possible to create a Monte Carlo algorithm that generates jets from initial partons, where the fragmentation functions obey Eqs.~(\ref{eq:DGLAPq}) and~(\ref{eq:DGLAPG}), as was done in many earlier works~\cite{Zapp:2009ud,Zapp:2013vla,Sjostrand:2006za}.
Here, we present a Monte Carlo algorithm where the individual steps of selecting momentum fractions and parton virtualities agree with other algorithm, e.g.~\cite{Zapp:2008gi}, while constraints in phase space, such as those for the case of VLEs given in Subsection~\ref{sec:vle}, differ (also, a fixed QCD coupling constant is used in our approach).
The jet particles are specified by their type (i.e.\ whether they are quarks or gluons) and their four-momenta. 
The algorithm starts from a given initial jet parton.
Then, for any jet parton $A$, it is determined whether a splitting occurs, and in such a case the splitting products $B$ and $C$ are obtained. This process is repeated for the splitting products $B$ and $C$, and for their possible products, etc., resulting in a parton cascade. 
Once no splitting occurs for any of the considered particles, the algorithm automatically stops. 
Then, it is still possible for particles to undergo the BDIM evolution inside the medium -- this is discussed later on.
The following few steps give the part of the algorithm that generates the splitting products $B$ and $C$ from the parton $A$ which is presumed to split:

\begin{enumerate}
    \item It is determined, whether the splitting produces a quark--antiquark pair.
    If the parton $A$ is a quark, it decays into a quark and a gluon, 
    otherwise a random number $R_t\in[0,1]$ is selected and compared to the ratio
    \begin{equation}
        p_t(Q_A^2,E_A)=\left[{1+\frac{\rho_{qg}(x_+(Q_A^2,E_A,R_{\rm max}),Q_A^2,E_A)}{\rho_{gg}(x_+(Q_A^2,E_A,R_{\rm max}),Q_A^2,E_A)}}\right]^{-1}\,.
    \end{equation}
    If $R_t\leq p_t(Q_A^2,E_A)$ a gluon-pair is produced, otherwise a quark--antiquark pair.
    \item The energy fraction $z_B=E_B/E_A$ is obtained by selecting a random number $R_z\in[0,1]$
    and solving the equation
    \begin{equation}
    R_z=\frac{\rho_{BA}(z_B,Q_A^2,E_A,R_{\rm max})}{\rho_{BA}(x_+(Q_A^2,E_A),Q_A^2,E_A)}\,.
    \end{equation}
    The energy fraction of the other particle $z_C=E_C/E_A$ is obtained from the energy conservation: $z_C=1-z_B$.

    \item The squared virtualities $Q_i^2$ ($i=B,\,C$) of the particles $B$ and $C$ are selected in such a way that they follow the probability distribution functions $p_i(Q_A^2, Q_i^2,E_A)$ given by 
    \begin{equation}
        p_i(Q_A^2, Q_i^2,E_A)=\frac{\partial}{\partial Q_i^2}\Delta_i(Q_A^2,Q_i^2,E_A)\,,
    \end{equation}
    i.e.\ obeying the CDFs of the Sudakov functions of Eqs.~(\ref{eq:Sudq}) and~(\ref{eq:SudG}).
    This selection process is done via a veto algorithm {(cf.~\cite{Sjostrand:2006za} for details). For the selection of the squared virtuality $Q_i^2$, the distributions 
    \begin{align}
\tilde{\Delta}_q(Q_1^2,Q_2^2)&=\exp\left[-\int_{Q_2^2}^{Q_1^2}\frac{dQ^2}{Q^2}\int_{\epsilon_{\rm VLE}}^{1-{\epsilon_{\rm VLE}}} dz \frac{\alpha_s}{2\pi}P_{qq}(z)\right]\,,\\
\tilde{\Delta}_g(Q_1^2,Q_2^2)&=\exp\bigg[-\int_{Q_2^2}^{Q_1^2}\frac{dQ^2}{Q^2}\int_{\epsilon_{\rm VLE}}^{1-{\epsilon_{\rm VLE}}} dz  \frac{\alpha_s}{2\pi} P_{gg}(z)
\nonumber\\&+N_F P_{qg}(z)\bigg]\,,
\end{align}
with the numerical parameter $\epsilon_{\rm VLE}$ given in Table~\ref{tab:simpar}, are used as CDFs. However, an additional rejection condition is applied, which ensures that the values selected for $Q_i^2$ obey the CDFs of Eqs.~(\ref{eq:Sudq}) and~(\ref{eq:SudG}).}
    \item The components of momenta of the particles $B$ and $C$ transverse to the momentum of particle $A$ are calculated as 
    \begin{align}
            k_{\perp BC}^2&=\frac{1}{\sqrt{1-\frac{Q_A^2}{E_A^2}}}\left[ z_B(1-z_B)Q_A^2-(1-z_B)Q_B^2-z_BQ_C^2 \right.\nonumber\\&\left.-\frac{1}{4E_A^2}\left(Q_A^4+Q_B^4 +Q_C^4 -2Q_A^2Q_B^2-2Q_A^2Q_C^2-2Q_B^2Q_C^2\right)\right]\,,
        \label{eq:kTsquared}    
    \end{align}
    If $k_{\perp\,BC}^2< 0$, steps 1 to 3 are repeated.
    \item An azimuthal angle $\phi$ of the transverse momentum of particle $B$ with regard to the momentum of the particle $A$ is selected randomly from the uniform distribution in the range $[0,2\pi]$. The corresponding angle for the particle $C$ is obtained as $\phi_C=\phi+\pi$. Thus, azimuthal correlations between the subsequent splittings are neglected in this algorithm. 
    \item Four-momenta of the particles $B$ and $C$ are calculated from the four-momentum of the particle $A$ and from the previously obtained variables $z_B$, $Q_B$, $Q_C$, $k_{\perp BC}^2$ and $\phi$, as described in a more detail in \ref{sec:kin} below.
\end{enumerate}
After the splitting of the parton $A$ into the partons $B$ and $C$, the four-momenta of the three particles are used to evaluate phase-space constraints that are outlined below in \ref{sec:stopcond}.
If the phase-space constraints are fulfilled, the partons $B$ and $C$ also split and the steps 1 to 6 in the list above are repeated for the partons $B$ and $C$. Otherwise, it is presumed that the splitting does not take place and the particle $A$ is considered as the final particle that does not undergo further evolution in the vacuum. The jet evolution by VLEs stops once all the generated particles are the final particles.

\subsection{Kinematical descriptions of particle four-momenta}\label{sec:kin}
This subsection describes how the jet-particle four-momenta are obtained (in the case of jets evolving via VLEs) from the variables $z_B$, $Q_B$, $Q_C$, $k_{\perp BC}^2$ and $\phi$ that were selected in step 6 of the algorithm outlined in the previous subsection. To this end let us refer to the momentum of the initial particle of the parton cascade as $p_{ini}$ and define as a jet-frame a reference frame, where $p_{ini}$ is given as
\begin{equation}
    p_{ini,\,jet}=\left(E,0,0,\sqrt{E^2-Q_{ini}^2}\right)\,.
    \label{eq:jet-frame}
\end{equation}

In \tmdice, the four-momentum of any jet-particle is obtained in a reference frame given by the jet-frame 
in the following way: For any jet-particle $A$ that splits into a pair of particles $B$ and $C$, the four-momentum of particle $A$ in the jet-frame $p_{A,\,jet}$ is known before the splitting into particles $B$ and $C$. After the splitting the four momenta of particles $A$, $B$ and $C$ are obtained in a reference frame where the z-axis is the same as the three-momentum of particle $A$:
%
\begin{align}
    p_{A,\,\mathrm{local}}&=\left(E_A,0,0,\sqrt{E_A^2-Q_A^2}\,\right)\,,\label{eq:palocal}\\
    p_{B,\,\mathrm{local}}&=\bigg(z_B\,E_A,\,k_{\perp BC}\cos\phi,\,k_{\perp BC}\sin\phi,\nonumber\\&\hspace{6mm}\sqrt{z_B^2E_A^2-k_{\perp BC}^2-Q_B^2}\,\bigg)\,,\label{eq:pblocal}\\
    p_{C,\,\mathrm{local}}&=\bigg(z_CE_A ,-k_{\perp BC}\cos\phi,-k_{\perp BC}\sin\phi,\nonumber\\&\hspace{6mm}\sqrt{z_C^2E_A^2-k_{\perp BC}^2-Q_C^2}\,\bigg)\,,\label{eq:pclocal}
\end{align}
with $z_c=1-z_b$.

The parton momenta $p_{B,\,\mathrm{local}}$ and $p_{C,\,\mathrm{local}}$ are then obtained in the jet-frame
as 
\begin{align}
    p_{B,\,jet}&=\hat{R}(\theta_{A},\phi_{A})\,p_{B,\,lab}\,,\\
    p_{C,\,jet}&=\hat{R}(\theta_{A},\phi_{A})\,p_{C,\,lab}\,,
    \label{eq:pjet}
\end{align}
by application of the following rotation matrix:
\begin{equation}
    \hat{R}(\theta_{A},\phi_{A})=\left(
\begin{array}{cccc}
 1 & 0 & 0 & 0 \\
 0 & \cos (\theta_{A} ) \cos (\phi_{A} ) & -\sin (\phi_{A} ) & \sin (\theta_{A} ) \cos (\phi_{A} ) \\
 0 & \cos (\theta_{A} ) \sin (\phi_{A} ) & \cos (\phi_{A} ) & \sin (\theta_{A} ) \sin (\phi_{A} ) \\
 0 & -\sin (\theta_{A} ) & 0 & \cos (\theta_{A} ) \\
\end{array}
\right)\,,
\end{equation}
where
\begin{align}
    \sin(\theta_{A})&=\frac{\sqrt{p_{A,\, jet \,x}^2+p_{A,\, jet \,y}^2}}{\sqrt{p_{A,\, jet\, x}^2+p_{A,\, jet\,y}^2+p_{A,\, jet \,z}^2}}\,,\\
    \sin(\phi_{A})&=\frac{p_{A,\, jet \,y}}{\sqrt{p_{A,\, jet \,x}^2+p_{A,\, jet \,y}^2}}\,.
\end{align}
Note that the above rotation differs for every parton branching due to the different angles $\theta_{A}$ and $\phi_{A}$.

Furthermore, in case the \tmdice~ and \katie~ algorithms are combined, all the jet-particle momenta can be rotated into the lab-frame of the collision.
%
%
The four-momentum of the initial jet particle in the lab frame $p_{ini,\,\mathrm{LAB}}$ is already known,
 and thus the rotation 
 that transforms $p_{ini,\,\mathrm{jet}}$ into $p_{ini,\,\mathrm{LAB}}$. 
 Therefore, one can obtain the four-momentum of every jet particle $i$ as
\begin{align}
    p_{i,\,\mathrm{LAB}}&=\hat{R}(\theta_{jet},\phi_{jet})\,p_{i,\,\mathrm{jet}}\,,
    \label{eq:plab}
\end{align}
with
\begin{align}
    \sin(\theta_{jet})&=\frac{\sqrt{p_{ini,\, \mathrm{LAB} \,x}^2+p_{ini,\, \mathrm{LAB} \,y}^2}}{\sqrt{p_{ini,\, \mathrm{LAB}\, x}^2+p_{ini,\, lab \,y}^2+p_{ini,\, \mathrm{LAB} \,z}^2}}\,,\\
    \sin(\phi_{jet})&=\frac{p_{ini,\, \mathrm{LAB} \,y}}{\sqrt{p_{ini,\, \mathrm{LAB} \,x}^2+p_{ini,\, \mathrm{LAB} \,y}^2}}\,.
\end{align}
%

\subsection{Phase-space constraints of VLEs and their implementations}\label{sec:stopcond}

In \tmdice, after every branching of a parton $A$ into partons $B$ and $C$ due to VLEs is simulated, it is verified whether that branching fulfills certain phase-space conditions. 
Only the parton branchings that fulfill the phase-space constraints are considered to contribute to the jet-evolution. If a parton branching violates the phase-space constraints, the parton $A$ is considered as the final particle that does not undergo further evolution by VLEs.  However, the parton $A$ may still undergo further evolution in a medium by scatterings and medium-induced emissions.
The phase-space constraints are: 
\begin{itemize}
\item It is tested whether Eq.~(\ref{eq:stopcoh}) holds for the particle $A$.
\item It is tested whether Eq.~(\ref{eq:stoplength}) holds for the particle $A$.
\item It is tested whether
\begin{equation}
k_{\perp\,BC}>k_{\perp \,\,\mathrm{min}}\,,
\label{eq:stopcond3}
\end{equation}
where $k_{\perp\,\,\mathrm{min}}$ is a manually set threshold value.
\end{itemize}

A particle that violates any of the three conditions described above 
via Eqs.~(\ref{eq:stopcoh}),~(\ref{eq:stoplength}) and (\ref{eq:stopcond3}) does no longer evolve via VLEs. However, it is possible that these final particles from VLEs still evolve via the medium-induced radiation and the scattering in the medium. 
To this end, all the final particles from VLEs above a certain threshold value $x_{\mathrm{min\,VLE}}$ of the energy fraction $x$ are considered the initial particles of the parton cascades formed via the medium-induced radiation and the scattering in the medium.

It is also possible that the final particles of the VLE evolution are still off-mass-shell. 
Then, after evolving in the medium, parton emissions outside the medium may occur as a possible third stage of the jet evolution~\cite{Caucal:2018dla}.
In the current work, we neglect such a third stage and set virtualites of the final particles of the VLE evolution to $0$. As a consequence of the vanishing parton virtualities, the particle four-momenta need to be reevaluated in the following way:
If the parton $A$ splits into the partons $B$ and $C$,  and the parton $B$ is the final particle from VLEs, then we set $Q_B=0$ and recalculate $k_T$ according to Eq.~(\ref{eq:kTsquared}). Then, Eqs.~(\ref{eq:pblocal})--(\ref{eq:plab}) are evaluated again, yielding new values for the four-momenta of the particles $B$ and $C$.
\subsection{Verifications of the algorithm for VLEs}
In order to verify that the part of TMDICE for VLEs correctly reproduces how the jet-particle energies evolve, we compare quark multiplicity distributions at a specific virtuality scale $Q_{\rm min}$ obtained from an initial quark of a fixed maximal virtuality $Q_{\rm max}$ with the corresponding results for the quark fragmentation functions obtained by numerically solving the same DGLAP equation. 
To this end, the DGLAP equation for quarks (i.e.\ neglecting emissions off gluons) was evolved from the initial condition of a fragmentation function $D_0(x,Q^2_{max})$ given at the maximal virtuality $Q_{max}$ down to some minimal virtuality scale $Q_{min}$ using the Euler method for solving of the first order differential equation in time.

\begin{figure}[!ht]
\centering
\includegraphics[scale=0.5]{"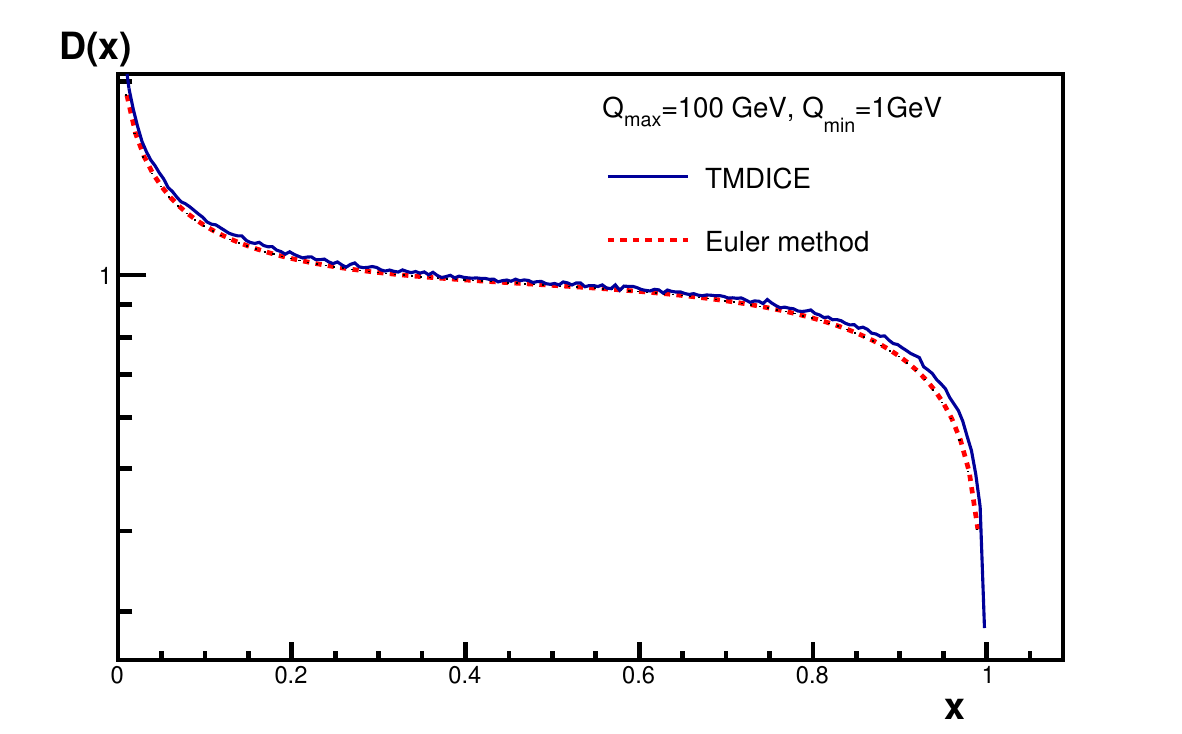"}
\caption{Quark fragmentation functions at $Q_{min}=1$~GeV obtained from the initial distributions $D_0(x,Q^2_{max})$ at $Q_{max}=100\,$GeV both via either the Euler method or the Monte Carlo algorithm of \tmdice\ as indicated in the legend. }
\label{fig:Eulercomp}
\end{figure}
Fig.~\ref{fig:Eulercomp} compares the fragmentation functions obtained by the Euler method and the Monte Carlo algorithm of \tmdice\ at $Q_{min}=1$~GeV from a fragmentation function 
\begin{equation}
    D_0(x,Q^2_{max})=\sqrt{\frac{2}{\pi \sigma^2}}\, {\rm e}^{-\frac{(1-x)^2}{2\sigma^2}}\,,
\end{equation}
with $\sigma=10^{-3}$. As can be seen, a good agreement between the two solutions has been obtained.
We point out that this comparison only works, if the condition 
that the momentum components of the emitted particles transverse to the momentum of the decaying particle $k_T>0$ has to fulfill after each splitting is deactivated in the Monte Carlo algorithm. Thus, the momentum fraction $x$ in both the Euler method and in the Monte Carlo algorithm are constrained in the same way as $x_{\rm min}<x<1-x_{\rm min}$.

\begin{figure}[!ht]
\includegraphics[scale=0.45]{"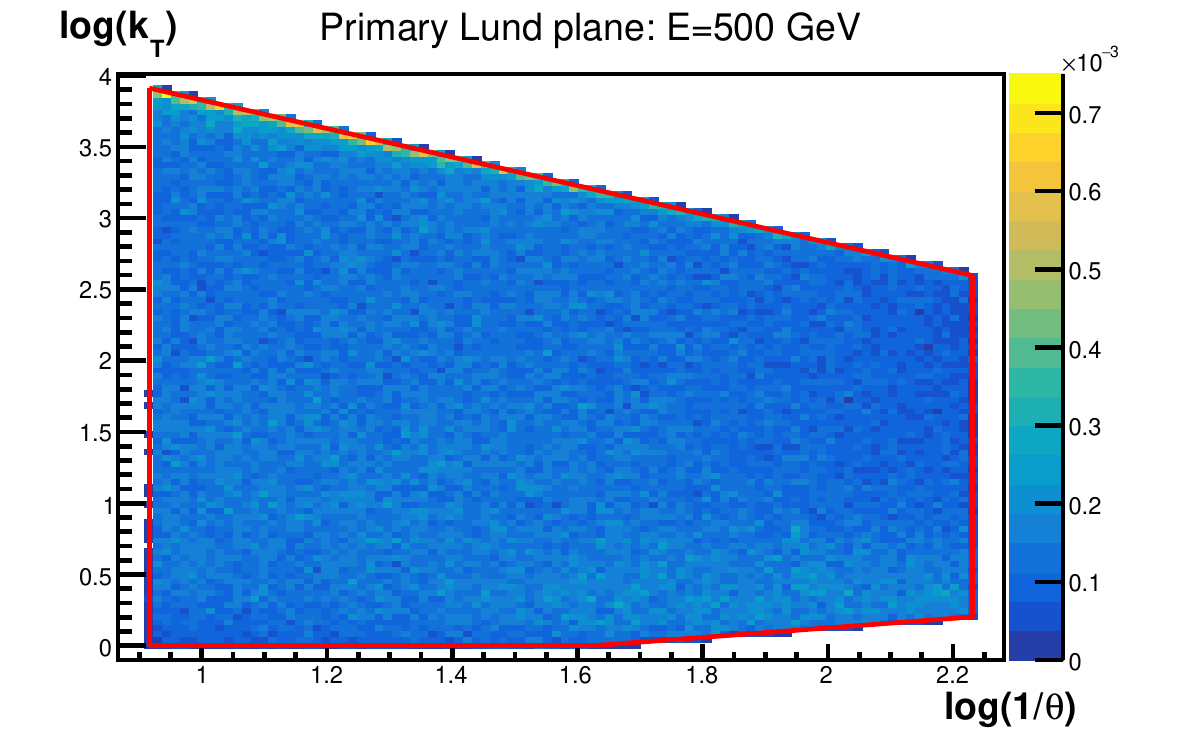"}
\includegraphics[scale=0.45]{"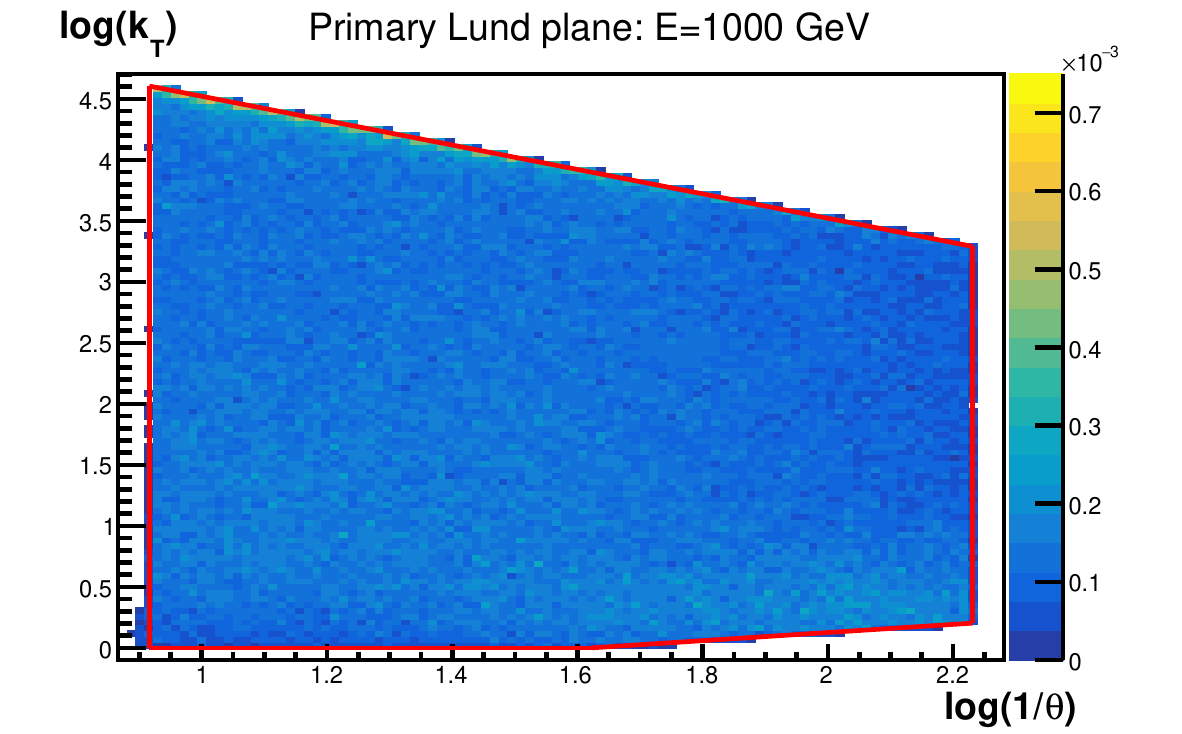"}
\caption{Primary Lund-planes for jets initiated by a quark of $E=500\,$GeV (top) and $E=1000$\,GeV (bottom). The red lines are the phase-space boundaries for VLEs as described in the text.  }
\label{fig:lundtest}
\end{figure}
Additionally, we show in Fig.~\ref{fig:lundtest} how the primary Lund planes are filled for quark initiated jets that undergo VLEs. 
The initial quarks are given either via $E=500\,$GeV, $Q_{max}=200\,$GeV and $R=0.4$ (left panel) or $E=1000\,$GeV, $Q_{max}=400\,$GeV and $R=0.4$ (right panel). The medium parameters are 
\begin{align}
    \hat{q}=1.5~{\rm GeV}^2/{\rm fm}\,, && L=3{\rm fm}\,.
\end{align}
The evolution via VLEs is constrained by the phase space conditions: 
\begin{itemize}
    \item $t_{\rm br}<t_{\rm decoh}$,
    \item $t_{\rm br}<L$,
    \item $k_T>k_{T\,, min}$,
\end{itemize}
with $k_{T\,, min}=1$~GeV here. As can be seen, the VLE parton branchings obey the phase-space constraints that are highlighted in the figure as red lines. These are: 
\begin{itemize}
\item $\ln k_{T} =\ln{k_{T,\,min}}=0$,
\item for $t_{\rm br}=t_{\rm decoh}$: $\ln k_T =-\frac{1}{3}\ln\frac{3\hat{q}}{2}+\frac{1}{3}\ln\frac{1}{\theta}$,
\item $\ln\frac{1}{\theta}=\ln\frac{1}{\theta_c},\;$ with $\theta_c=\sqrt{\frac{12}{\hat{q}L^3}}$.
\end{itemize}
The distributions in Fig.~\ref{fig:lundtest} are rather flat, but not completely uniform. The uniform distribution is not to be expected, since the full LO DGLAP splitting functions were used as evolution kernels, and not only the leading-log approximation in $x$.
\bibliographystyle{elsarticle-num} 
\bibliography{main}
\end{document}

%% file: definitions.tex
\def\sst{\scriptscriptstyle}

\long\def\comment#1{ }

\definecolor{kkcolor}{rgb}{0,1,0}

\def\raa{{R_{AA}}}

\def\pt{p_{\sst T}}
\def\pT{p_{\sst T}}

\newcommand{\beq}{\begin{eqnarray}}
\newcommand{\eeq}{\end{eqnarray}}
\newcommand{\be}{\begin{eqnarray*}}
\newcommand{\ee}{\end{eqnarray*}}
\newcommand{\rmd}{{\rm d}}

\newcommand{\rme}{{\rm e}}

\def\qg{{\rm qg}}

\newcommand{\nn}{\nonumber\\ }

\newcommand{\bem}{\begin{multline}}
\newcommand{\eem}{\end{multline}}
\newcommand{\beg}{\begin{gather}}
\newcommand{\eeg}{\end{gather}}

\newcommand{\ben}{\begin{eqnarray*}}
\newcommand{\een}{\end{eqnarray*}}

\newcommand{\secn}[1]{Section~1}
\newcommand{\appn}[1]{Appendix~1}
\long\def\comment#1{ }

\def\and{\quad\text{and}\quad}

\def\0{{\boldsymbol 0}}

\usepackage{textcomp}

\newcommand{\katie}{{\sf KATIE}}

\newcommand{\tmdice}{{\sf TMDICE}}

\newcommand {\photonjet}{$\gamma$-jet}
